\documentclass[
 aps,pra,11pt,
 twoside,
 amsmath,
 amssymb,
 balancelastpage,
 eqsecnum, graphicx
 ]{revtex4-1}
\usepackage{siunitx}
\usepackage{graphics} 
\usepackage[]{natbib}
\usepackage{color}
\usepackage{hyperref}		
\usepackage[dvipdfmx]{graphicx} 
\usepackage{bmpsize}
\usepackage{array}

\begin{document}


\title{Cold ion chemistry} 


\author{Dongdong Zhang}
\email[]{dongdong.zhang@unibas.ch}
\affiliation{Department of Chemistry, University of Basel, Klingelbergstrasse 80, 4056 Basel,Switzerland}

\author{Stefan Willitsch}
\email[]{stefan.willitsch@unibas.ch}
\affiliation{Department of Chemistry, University of Basel, Klingelbergstrasse 80, 4056 Basel,Switzerland}


\date{\today}

\begin{abstract}
Studying chemical reactions at very low temperatures is of importance for the understanding of fundamental physical and chemical processes. At very low energies, collisions are dominated by only a few partial waves. Thus, studies in this regime allow the characterization of quantum effects which depend on the collisional angular momentum, e.g., reactive scattering resonances and tunneling through centrifugal barriers. Additionally, the dynamics of ultralow-energy collisions is dominated by long-range interactions, i.e., "universal" chemical forces. Thus, experiments in this domain probe the details of intermolecular interactions which is of general relevance for the understanding of chemical processes. Moreover, the increasingly more precise experiments provide valuable data for benchmarking theoretical models and quantum-chemical calculations.

Studies of cold ion-neutral reactions in the laboratory rely on techniques for the generation of cold ions and cold neutrals. Over the last decades, the technology in this domain has made impressive progress. Many of these techniques do not only allow the translational cooling of the molecules, but also a precise preparation of their internal quantum state. Thus, by an accurate preparation of both the energy and state of the reaction partners as well as by the application of external electric, magnetic or optical fields, prospects open up to study and control chemical processes at an unprecedented level of accuracy. In this chapter, we will review salient theoretical concepts and recent experimental developments for studies of ion-molecule reactions at low temperatures. We first discuss the theoretical background for the description of low-energy ion-molecule reactions. Next we present key experimental methods for laboratory studies of cold ion-molecule reactions. The chapter concludes with a review of illustrative results and an outlook on future directions.
\end{abstract}


\maketitle 
\tableofcontents
\newpage


\section{Introduction}\label{se:intro}

Studies of ion-molecule reactions at low temperatures have received considerable attention since the early years of the 21st century. One of the major motivations for experiments in this domain is that low-temperature rate constants are required for modeling the chemistry of interstellar clouds in which temperatures on the order of 10~K prevail. An important class of reactions in the interstellar medium are ion-molecule processes with no activation barriers which usually have large reaction rates at low temperatures \cite{smith06a,smith06b,
snow08a,bodo09a,smith11a,smith11b,herbst11a}. Such reactions can lead to the formation of larger molecules and are at the root of the chemical diversity of our universe. 

Studying chemical reactions at very low temperatures is also of importance for the understanding of fundamental physical and chemical processes. At very low energies, collisions are dominated by only a few (in the extreme case even only one) partial waves, i.e., wave functions with a well-defined collisional angular momentum quantum number. Thus, studies in this regime allow the characterization of quantum effects which depend on the collisional angular momentum, e.g., reactive scattering resonances and tunneling through centrifugal barriers \cite{weck06a,bell09b,brouard12a}. While this domain has already been reached in collisions between neutrals \cite{ospelkaus10b, henson12a, chefdeville13a}, it remains one of the outstanding challenges of the future to access this regime in ion-neutral experiments \cite{cetina12a}. Additionally, the dynamics of ultralow-energy collisions is dominated by long-range interactions, i.e., ''universal'' chemical forces which do not depend on short-range interactions specific to the system. Thus, experiments in this domain probe the details of intermolecular interactions which is of general relevance for the understanding of chemical processes. Moreover, the increasingly more precise experiments provide valuable data for benchmarking theoretical models and quantum-chemical calculations.

Studies of cold ion-neutral reactions in the laboratory rely on techniques for the generation of cold ions and cold neutrals. Over the last decades, the technology in this domain has made impressive progress. A large variety of molecular species can now be prepared at cold ($<$ 1 \si{\kelvin}) or even ultracold ($<$ 1 \si{\milli\kelvin}) translational temperatures. Many of these techniques do not only allow the translational cooling of the molecules, but also a precise preparation of their internal quantum state  \cite{bethlem03a,hutson06a, koehler06a,willitsch08b,meerakker08a, krems09a, schnell09a,bell09b, carr09a, dulieu09a, chin10a, hutzler12a, ulmanis12a, willitsch12a, stuhl14a, meerakker12a, jankunas15a, heazlewood15a}. Thus, by an accurate preparation of both the energy and state of the reaction partners as well as by the application of external electric, magnetic or optical fields, prospects open up to study and control chemical processes at an unprecedented level of accuracy \cite{chang15a,willitsch16a}.

In this chapter, we will review salient theoretical concepts and recent experimental developments for studies of ion-molecule reactions at low temperatures. In Sec. \ref{se:theory}, the theoretical background for the description of low-energy ion-molecule reactions is discussed. In Sec. \ref{se:methods}, we will present key experimental methods for laboratory studies of cold ion-molecule reactions. The chapter concludes with a review of illustrative results given in Sec. \ref{se:experiment} and an outlook on future directions in Sec. \ref{se:conclusion}.

\section{Theory of ion-molecule collisions and reactions}\label{se:theory}

Collisions between molecules can be categorized as elastic (only kinetic energy is exchanged between the collision partners), inelastic (kinetic and internal energy is exchanged) and reactive (the chemical structure of the collision partners is changed). For a detailed, general introduction to the quantum theory of molecular collisions, we refer the reader to Ref. \cite{brouard12a,balint-kurti15a}, and to Chapter XX by G. Qu\'{e}m\'{e}ner in this book. Most of the basic quantities defined below are formulated within the specific context of ion-atom collisions.
\subsection{Elastic collisions}\label{ss:elastic}
We start with an outline of the quantum theory of elastic collisions which represent the simplest form of collisional processes. In the context of the present chapter, elastic collisions are of relevance for the buffer-gas and sympathetic cooling of ions by cold atoms, see Secs. \ref{ss:buffergas} and \ref{ss:lasercooling}.

In quantum collision theory, the elastic collision cross section can be formulated as \cite{connor73a,brouard12a,balint-kurti15a,cote00a}: 
\begin{equation}\label{eq:crosssection}
\sigma_{el}=\dfrac{4\pi}{k^{2}}\sum_{l=0}^{\infty}(2l+1)sin^{2}(\eta_{l}),
\end{equation}
where $k=\sqrt{2\mu_{R} E_{col}}/\hbar$ is the collision wave number, $\mu_{R}$ is the reduced mass of the colliding particles, $E_{col}$ is their relative collision energy in the center-of-mass system, $l$ is the collisional angular momentum quantum number and $\eta_{l}$ is the phase shift of the $l$th partial-wave. A scattering wave function with a well-defined angular momentum quantum number $l$ is referred to as a partial wave. $\eta_{l}$ is the phase shift imprinted on the wavefunction by the interaction between the particles. The problem now reduces to the determination of this scattering phase. 
The phase shift $\eta_{l}$ can be determined from solving the radial scattering Schr\"odinger equation at a given collision energy $E_{col}$ for the wavefunction $y_{E_{col},l}(R)$
\begin{equation}\label{eq:partialwaveequation}
\bigg[\dfrac{d^{2}}{dR^{2}}+k^{2}-\dfrac{l(l+1)}{R^{2}}-\dfrac{2\mu_{R}}{\hbar ^{2}}V(R)\bigg]y_{E_{col},l}(R)=0,
\end{equation}
where $R$ is the distance between the centers-of-mass of the collision partners  and $V(R)$ is the interaction potential. The asymptotic solution of $y_{E_{col},l}$($R\rightarrow\infty$) is given by
\begin{equation}\label{eq:phaseshift}
y_{E_{col},l}(R)\sim sin\bigg[kR-\dfrac{l\pi}{2}+\eta_{l}(k)\bigg].
\end{equation}

For collisions of ions with structureless neutral particles, the long-range potential is dominated by the charge-induced dipole interaction, sometimes also referred to as polarization or Langevin interaction. In this case, $V(R)=-C_{4}/R^{4}$ (see Sec. \ref{ss:capture}). Ion-molecule collisions at very low energies, e.g., in the sub \si{\micro\kelvin} regime for Na$^{+}$ + Na \cite{cote00a}, are dominated by $s$ waves, i.e., $l=0$. A modified effective-range formula can be used to determine the phase shift \cite{malley61a,cote00a} 
\begin{equation}\label{eq:effectiverange}
kcot(\eta_{0}(k)) \simeq -\dfrac{1}{a_{0}} + \dfrac{\pi}{3}\dfrac{2\mu_{R}}{\hbar^{2}}\dfrac{C_{4}}{a_{0}^{2}}k + \dfrac{4}{3}\dfrac{2\mu_{R}}{\hbar^{2}}\dfrac{C_{4}}{a_{0}}k^{2}ln\bigg(\dfrac{k}{4}\sqrt{•\dfrac{2\mu_{R}}{\hbar^{4}}C_{4}}\bigg).
\end{equation}
Here, $C_{4}$ denotes the charge-induced dipole interaction coefficient (see Sec. \ref{ss:capture}) and $a_{0}$ is the $s$-wave scattering length which is a measure for the range of the potential \cite{balint-kurti15a, brouard12a}.
At higher collision energies, usually several partial waves contribute to the collisions so that it is often sufficiently accurate to evaluate the phase shifts using the Wigner-Kramers-Brillouin (WKB) approximation \cite{langer37a,dalgarno58a}
\begin{equation}\label{eq:JWKB}
\eta_{l}\simeq\int_{R_{0}}^{\infty}\bigg[k^{2}-\dfrac{2\mu_{R} V(R)}{\hbar^{2}}-\dfrac{l(l+1)}{R^{2}}\bigg]^{1/2}dR-\int_{R_{0}^{'}}^{\infty}\bigg[k^{2}-\dfrac{l(l+1)}{R^{2}}\bigg]^{1/2}dR.
\end{equation}
The lower limits $R_{0}$ and $R_{0}^{'}$ correspond to the outermost zeros of the relevant integrands. 
For the general case, it is not possible to evaluate Eq. (\ref{eq:JWKB}) analytically. However for large values of $l$, it has been shown that Eq. (\ref{eq:JWKB}) can be approximated by \cite{massey34a,dalgarno58a}
\begin{equation}\label{eq:classical}
\eta_{l}\simeq-\dfrac{\mu_{R}}{\hbar^{2}}\int_{R_{0}^{'}}^{\infty}\dfrac{V(R)}{\bigg[k^{2}-\dfrac{(l+1/2)^{2}}{R^{2}}\bigg]^{1/2}}dR.
\end{equation}
Considering the interaction potential $V(R)=-C_{4}/R^{4}$ (see Sec. \ref{ss:capture}) an analytical solution can be derived if the outer turning point $R_0$ falls in the asymptotic region \cite{cote00a}:
\begin{equation}\label{eq:solution}
\eta_{l}\simeq\dfrac{\pi \mu_{R}^{2} C_{4}}{4\hbar^{4}}\dfrac{E_{col}}{l^{3}}.
\end{equation}

Substituting Eq. (\ref{eq:solution}) into Eq. (\ref{eq:crosssection}) gives an analytical expression of the elastic cross section for collisions between ions and molecules with a polarization interaction. Further approximations can be made if numerous phase shifts contribute. At $l\equiv L$ where $L$ is large enough so that Eq. (\ref{eq:solution}) is valid and $\eta_l$ is small, $sin(\eta_{l})$ can be approximated by $\eta_{l}$. If additionally the phase shifts quickly become large as $l$ decreases below $L$, $sin^{2}(\eta_l)$ may be approximated by its averaged value of 1/2. This approximation gives the cross section as
\begin{equation}\label{eq:corsssection app}
\sigma_{el}(E_{col})=\dfrac{2\pi}{k^{2}}L^{2}[1+\eta_{L}^{2}].
\end{equation}
A reasonable choice for $L$ would be such that $\eta_{L}=\pi/4$, i.e., $sin^{2}(\eta_{L})=1/2$ \cite{cote00a}. Then,
\begin{equation}\label{eq:corsssection final}
\sigma_{el}(E_{col})=\pi\bigg(\dfrac{\mu_{R}C_{4}^{2}}{\hbar^{2}}\bigg)^{1/3}\bigg(1+\dfrac{\pi^{2}}{16}\bigg)E_{col}^{-1/3}.
\end{equation}

In the cold regime ($1~\si{\kelvin}>T>1~\si{\milli\kelvin}$) accessed in current experiments, this semiclassical approach is often sufficiently accurate to describe elastic collisions \cite{cote00a}. An alternative derivation of the elastic ion-neutral scattering cross section using the Eikonal approximation is outlined in Ref. \cite{hudson09a}. In many cases, the elastic scattering rate can also be approximated by the Langevin rate \cite{zipkes11a, chen14a, rouse15a}, see Sec. \ref{ss:capture}.


\subsection{Inelastic collisions}\label{ss:close-coupled}

Inelastic collisions play an important role in a variety of contexts, from buffer-gas cooling of the internal degrees of freedom (see Sec. \ref{ss:buffergas}) to energy transfer in astrochemical environments, see, e.g., Refs. \cite{hammami07a,gomez-carrasco14a}. The quantum-mechanical treatment of inelastic collisions is usually formulated within the framework of close-coupled scattering equations \cite{brouard12a,balint-kurti15a}. Here, we present a brief outline of the salient ideas. As an example, consider a rotationally inelastic collision process between an atom A and a diatomic molecule BC,  A + BC($j$) $\rightarrow$ A + BC($j'$), where $j$ and $j'$ are the rotational quantum number of the molecule. The Hamiltonian of the problem can be written as 
\begin{equation}\label{eq:cc:equ}
\hat{H} = -\dfrac{\hbar^{2}}{2\mu_{R}}\nabla_R^{2} + \hat{h} + V,
\end{equation}
where $\mu_{R}$ = $m_{A}m_{BC}/(m_{A}+m_{BC})$ is the reduced mass of the colliding particles, $\hat{h}$ is the Hamiltonian of the diatomic molecule BC 
\begin{equation}\label{eq:cc:h}
\hat{h} = -\dfrac{\hbar^{2}}{2\mu_{r}}\nabla_r^{2} + w(r),
\end{equation}
$\mu_{r}$ is the reduced mass of the molecule BC, $w(r)$ is the intermolecular potential of BC, and $V$ is the interaction potential between A and BC.
Inserting Eq. (\ref{eq:cc:h}) into Eq. (\ref{eq:cc:equ}) and absorbing $w(r)$ into $V(R)$, we get
\begin{equation}\label{eq:cc:totalH}
\hat{H} = -\dfrac{\hbar^{2}}{2\mu_{R}}\dfrac{1}{R}\dfrac{\partial^{2}}{\partial R^{2}}R + \dfrac{\hat{\Vec{l}}^{2}}{2\mu_{R}R^{2}} -\dfrac{\hbar^{2}}{2\mu_{r}}\dfrac{1}{r}\dfrac{\partial^{2}}{\partial r^{2}}r + \dfrac{\hat{\Vec{j}}^{2}}{2\mu_{r}r^{2}} + V(R,r),
\end{equation}
where $R$ is the distance between A and the center-of-mass of BC, $\hat{\Vec{l}}$ and $\hat{\Vec{j}}$ are the collisional and molecular rotational angular momentum operators, respectively, and $r$ is the distance between B and C. The total Hamiltonian $\hat{H}$ commutes with the total angular momentum operator defined as \cite{brouard12a}
\begin{equation}\label{eq:cc:angular}
\hat{\vec{J}} = \hat{\vec{j}} + \hat{\vec{l}},
\end{equation}
and its projection on the $z$ axis of the spaced-fixed frame
\begin{equation}\label{eq:cc:angularprojection}
\hat{J_{z}} = \hat{j_{z}} + \hat{l_{z}}.
\end{equation}
The angular part of the scattering wavefunction can be formulated by the ansatz
\begin{equation}\label{eq:cc:angularsolution}
\psi_{jl}^{JM_{J}} = \sum_{m_{j} m_{l}} \varphi_{lm_{l}}\varphi_{jm_{j}}\langle jm_{j},lm_{l}\mid JM \rangle,
\end{equation}
where $\langle jm_{j},lm_{l}\mid JM \rangle$ is a Clebsch-Gordan coefficient and the angular functions $\varphi_{lm_{l}}$ and $\varphi_{jm_{j}}$ are eigenfunctions of the angular-momentum operators $\hat{\Vec{l}}^2$ and $\hat{\Vec{j}}^2$, respectively \cite{zare88a}.

From Eq. (\ref{eq:cc:angularsolution}), it is obvious that the $\psi_{jl}^{JM_{J}}$ are eigenfunctions of all four angular momentum operators $\hat{\vec{J}}^{2}$, $\hat{J_{z}}$, $\hat{\vec{j}}^{2}$ and $\hat{\vec{l}}^{2}$ with the  eigenvalues $\hbar^{2}J(J+1)$, $\hbar M_{J}$, $\hbar^{2}j(j+1)$ and $\hbar^{2}l(l+1)$ after normalization.

At long range, where $R\rightarrow\infty$, the interaction between particle A and BC is weak and can be neglected. The only term left in the interaction potential $V$ is then the internal potential $w(r)$ of the molecule BC. Let us now consider the molecular Hamiltonian $\hat{h}$ the eigenfunctions of which can be written as
\begin{equation}\label{eq:cc:molecule}
\Phi_{vjm_{j}}(\vec{r}) = \dfrac{1}{r}\phi_{vj}(r)\varphi_{jm_{j}}(\hat{r}),
\end{equation}
where $v$ denotes the vibrational quantum number, $\hat{r}$ the set of angular coordinates associated to $\vec{r}$, and $\phi_{vj}(r)$ are normalized functions satisfying the radial Schr\"odinger equation \cite{balint-kurti15a}
\begin{equation}\label{eq:cc:moleculeradial}
\big[-\dfrac{\hbar^{2}}{2\mu_{r}}\dfrac{d^{2}}{dr^{2}} + \dfrac{j(j+1)}{2\mu_{r}r^{2}} + w(r)\big]\phi_{vj}(r) = E_{vj}\phi_{vj}(r).
\end{equation}
Combining Eq. (\ref{eq:cc:molecule}) and Eq. (\ref{eq:cc:angularsolution}), the solution $\Psi^{JM_J}$ of the Schr\"odinger equation 
\begin{equation}\label{eq:cc:schrodingerequ}
\hat{H}\Psi^{JM_{J}} = E\Psi^{JM_{J}}
\end{equation}
can be expressed as 
\begin{equation}\label{eq:cc:schrodingersolution}
\Psi^{JM_{J}} = \dfrac{1}{Rr}\sum_{v'j'l'}\chi_{v'j'l'}^{JM}(R)\psi_{j'l'}^{JM_{J}}(\hat{R}, \hat{r})\phi_{v'j'}(r),
\end{equation}
where the $\chi_{v'j'l'}^{JM}(R)$ denote radial expansion coefficients. From Eqs. (\ref{eq:cc:schrodingerequ}) and (\ref{eq:cc:schrodingersolution}), a set of "close-coupled equations" can be derived after integration over the angular coordinates \cite{balint-kurti15a}:
\begin{equation}\label{eq:cc:cc}
\big[\dfrac{d^{2}}{dR^{2}} + k_{vj}^{2} - \dfrac{l(l+1)}{R^{2}}\big]\chi_{vjl}^{JM}(R) = \dfrac{2\mu_{R}}{\hbar^{2}} \sum_{v'j'l'} V_{vjl,v'j'l'}^{JM}(R)\chi_{v'j'l'}^{JM}(R),
\end{equation}
with 
\begin{equation}\label{eq:cc:k}
k_{vj}^{2} = \dfrac{2\mu_{R}}{\hbar^{2}}[E_{col}-E_{vj}],
\end{equation}
and 
\begin{equation}\label{eq:cc:V}
V_{vjl,v'j'l'}^{JM}(R) = \int d\hat{R}\int d\hat{r}\int dr \phi_{vj}(r)^{\ast}\psi_{jl}^{JM}(\hat{R}, \hat{r})^{\ast}V\psi_{j'l'}^{JM}(\hat{R}, \hat{r}) \phi_{v'j'}(r).
\end{equation}
The notation $\hat{R}$ denotes the set of angular coordinates associated with $\Vec{R}$. Eq. (\ref{eq:cc:cc}) defines a set of coupled differential equations which needs to be solved to obtain the expansion coefficients $\chi_{vjl}^{JM}(R)$ of Eq. (\ref{eq:cc:schrodingersolution}) subject to appropriate boundary conditions \cite{balint-kurti15a}.

The close-coupled equations Eq. (\ref{eq:cc:cc}) must be solved numerically. Several approaches have been proposed, e.g., by finding piecewise analytic solutions or by the log-derivative method \cite{balint-kurti15a}. An alternative and elegant method to solve these equations for cold ion-atom collisions was recently formulated within the framework of multichannel quantum-defect theory (MQDT), see Refs. \cite{gao08a,idziaszek09a,raab09a,idziaszek11a,li04a} for details..


\subsection{Capture models}\label{ss:capture}

The approach outlined in the previous section can be expanded to the treatment of reactive collisions \cite{brouard12a, balint-kurti15a}. However, the integration of the scattering equations is computationally costly even for small systems. An alternative approach are classical or quasi-classical trajectory calculations, see, e.g., Refs. \cite{dugan67a,dugan72a,swamy82a,tong12a,li14a}. Moreover, a range of simplified approximate treatments has been developed for ion-molecule collisions which do not require elaborate numerical calculations and can also be applied to large systems.

\begin{figure}[!htb]
    \centering
    \resizebox{0.8\linewidth}{!}
    {\includegraphics{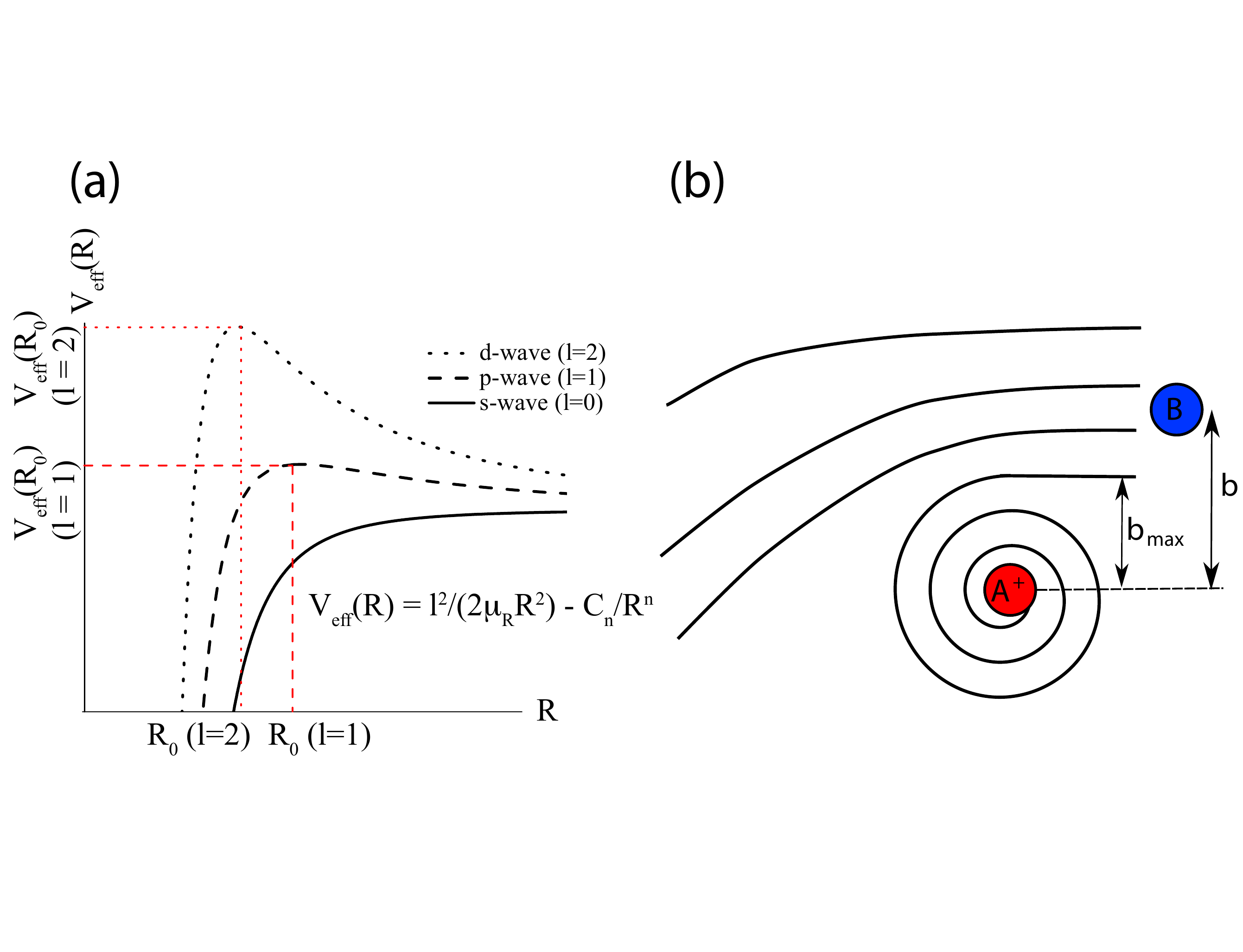}}
    \caption{(a). Effective potential $V_{eff}(R)$ as a function of the angular momentum $l$ and position $R_{0}$ of its maximum value. 
    (b). If the impact parameter is above the maximum value $b_{max}$, collisions only lead to weak deflection of the collision partners in glancing collisions. At impact parameters below this critical value, the collision partners enter into an inward-spiralling trajectory which results in a short-range collision.
}
    \label{fig:capture}
\end{figure}

As many ion-molecule reactions are barrierless, they are frequently modelled within the realm of captures theories. Capture models assume that the reaction rate is governed by long-range interactions leading to the capture of the collision partners, and that the short-range reaction probability is unity. In a classical picture, this means that all trajectories which pass the centrifugal potential barrier are inward-spiralling and lead to reaction \cite{gioumousis58a} (see Fig. \ref{fig:capture}).

Long-range intermolecular interactions are usually described by a series expansion in powers of $1/R$ \cite{buckingham67a}:
\begin{equation}\label{eq:capture:potential}
V(R)=\sum_{n}V_{n}(R), V_{n}(R)=-\dfrac{C_{n}}{R^{n}},
\end{equation}
where the $C_{n}>0$ (assuming attractive interaction) are the long-range coefficients.  In many cases, it is sufficient to only consider the leading term of the expansion Eq. (\ref{eq:capture:potential}).  For example, $n=2$ for a charge-dipole interaction and $n=4$ for a charge-induced dipole (polarization, Langevin) interaction. See Table \ref{tab:multipole} (for more details, see the chapter by Lepers and Dulieu in this book).

\begin{table}[!htb]
\caption{Long-range coefficients $C_n$ in atomic units. Explanation of symbols: molecular dipole moment $\mu_D$,  angle of orientation $\theta$ between ion-molecule distance vector and molecular axis, interparticle distance $R$, quadrupole moment $Q$, polarizability $\alpha$. }
\label{tab:multipole}
\begin{ruledtabular}
\begin{tabular}{ccc}
$n$&interaction type&Long-range coefficient $C_n$ \\\hline
2&charge-dipole& $-\mu_Dcos(\theta)$\\
3& charge-quadrupole&$Q(3cos^2(\theta)-1)/2$\\
4&charge-induced dipole& $-\alpha/2$\\
\end{tabular}
\end{ruledtabular}
\end{table}

If the particles collide at a non-zero impact parameter $b$, where $b$ is defined as the shortest distance between the collision partners in the absence of the interaction potential, then the collision has a non-vanishing angular momentum of magnitude $l=\mu_{R} v b$, with $v$ the collision velocity \cite{brouard12a}. In this case, a centrifugal-energy term $V_\text{cent}=l^2/(2\mu_{R} R^2)$ has to be added to the interaction potential yielding the effective potential of the collision
\begin{equation}\label{eq:capture:effectivepotential}
V_{eff}(R)=\dfrac{l^{2}}{2\mu_{R} R^{2}}-\dfrac{C_{n}}{R^{n}}.
\end{equation}

For $n>2$, the classical capture cross section can be derived as follows \cite{levine05a, brouard12a}. First, we find the position $R_0$ of the maximum in the effective potential (i.e., the position of the centrifugal barrier) by differentiating Eq. (\ref{eq:capture:effectivepotential}): (see Fig.~\ref{fig:capture} (a))
\begin{equation}\label{eq:capture:maxposition}
R_{0}=\bigg(\dfrac{n\mu_{R} C_{n}}{l^{2}}\bigg)^{\tfrac{1}{n-2}}.
\end{equation}
The value of the effective potential at $R_0$ is
\begin{equation}\label{eq:capture:maxeffectivepotential}
V_{eff}(R_{0})=\bigg(\dfrac{l^{2}}{\mu_{R}}\bigg)^{\tfrac{n}{n-2}}\dfrac{n-2}{2n}(nC_{n})^{-\tfrac{2}{n-2}}.
\end{equation} 
The trajectory is inward-spiralling and thus reactive if the collision energy $E_{col}$ exceeds the maximum effective potential energy $V_{eff}(R_{0})$. This results in a maximum value for $l$:
\begin{equation}\label{eq:capture:maxangular}
l_{max}=(\mu_{R} n)^{1/2}(C_{n})^{\tfrac{1}{n}}\bigg(\dfrac{2E_{col}}{n-2}\bigg)^{\tfrac{n-2}{2n}},
\end{equation} 
which corresponds to a maximum impact parameter (see Fig.~\ref{fig:capture} (b))
\begin{equation}\label{eq:capture:maximpact}
b_{max}=\dfrac{l_{max}}{\mu_{R} v},
\end{equation} 
and a cross section
\begin{equation}\label{eq:capture:crosssection}
\sigma(E_{col})=\pi b_{max}^{2}=\dfrac{\pi}{2}n\bigg(\dfrac{2}{n-2}\bigg)^{\tfrac{n-2}{n}}\bigg(\dfrac{C_{n}}{E_{col}}\bigg)^{\tfrac{2}{n}}.
\end{equation} 
Assuming thermal equilibrium, the thermally averaged rate constant is given by
\begin{equation}\label{eq:capture:rate}
k(T)=\int_{0}^{\infty}vf(v)\sigma(v)dv,
\end{equation} 
where $f(v)$ is the Maxwell-Boltzmann speed distribution \cite{brouard12a} 
\begin{equation}\label{eq:capture:vdistribution}
f(v)=4\pi\bigg(\dfrac{\mu_{R}}{2\pi k_{B}T}\bigg)^{3/2}v^{2}e^{-\tfrac{\mu_{R}v^{2}}{2k_{B}T}},
\end{equation} 
where $k_{B}$ and $T$ are the Boltzmann constant and temperature, respectively. Substituting Eq. (\ref{eq:capture:vdistribution}) into Eq. (\ref{eq:capture:rate}), the thermal rate coefficient is obtained as
\begin{equation}\label{eq:capture:rateT}
k(T)=\sqrt{\dfrac{2\pi}{\mu_{R}}}n\bigg(\dfrac{2}{n-2}\bigg)^{\tfrac{n-2}{n}}(C_{n})^{2/n}(k_{B}T)^{\tfrac{n-4}{2n}}\Gamma(2-\dfrac{2}{n}),
\end{equation} 
where $\Gamma$ is the mathematical Gamma function \cite{brouard12a}.
For instance, in the case $n = 2$, one finds $l_{max}$ = $\sqrt{2\mu_{R} C_{2}}$. Inserting it into Eq. (\ref{eq:capture:maximpact}) and Eq. (\ref{eq:capture:crosssection}), the cross section is obtained as
$\sigma = \pi C_{2}/E_{col}$.

In Table~\ref{tab:classRateSummary}, classical capture cross sections and thermal reaction rate constants for different long-range potentials are summarized. The result for $n=4$ corresponds to the important case of the capture of an ion by a structureless neutral particle \cite{gioumousis58a}, e.g., closed-shell atoms and, often in good approximation, also non-polar molecules (see Sec. \ref{ss:molbeam}). The corresponding expression for the rate constant is often also referred to as the "Langevin capture rate constant". It is a special case for which classical capture theory predicts a rate constant which does not depend on temperature. 

\begin{table}[!htb]
\caption{Cross sections $\sigma_{n}(E_{col})$ and thermal reaction rate constants $k_{n}(T)$ for different long-range potentials \cite{brouard12a}}
\label{tab:classRateSummary}
\begin{ruledtabular}
\begin{tabular}{cccc}
$n$&interaction type&$\sigma_{n}(E_{col})$&$k_{n}(T)$\\
\hline
2&charge-dipole&$\pi C_{2}/E_{col}$&$2\sqrt{2\pi/\mu_{R}}C_{2}(k_{B}T)^{-1/2}$\\
3&dipole-dipole&$3\pi(C_{3}/(2E_{col}))^{2/3}$&$4\sqrt{\pi /(3\mu_{R} )}\Gamma(1/3)(C_{3})^{2/3}(k_{B}T)^{-1/6}$\\
4&charge-induced dipole&$2\pi\sqrt{C_{4}/E_{col}}$&$2\pi\sqrt{2C_{4}/\mu_{R}}$\\
5&&$5\pi2^{1/2}3^{-3/2}(C_{5}/E_{col})^{2/5}$&$5\sqrt{2\pi/\mu_{R}}(2/3)^{3/2}(C_{5})^{2/5}(k_{B}T)^{1/10}\Gamma(3/5)$\\
6&dispersion&$3\pi(1/2)^{2/3}(C_{6}/E_{col})^{1/3}$ & $2^{11/6}\Gamma(2/3)\sqrt{\pi/\mu_{R}}(C_{6})^{1/3}(k_{B}T)^{1/6}$\\
\end{tabular}
\end{ruledtabular}
\end{table}

Capture models can also be developed on the basis of quantum scattering theory. To derive an expression for quantum capture, we formulate the integral cross section as a sum of partial cross sections:

\begin{equation}\label{eq:capture:DCSICS}
\sigma=\sum_{l=0}^{\infty}\sigma_{l} 
\end{equation}
where $\sigma_{l}$ is given by (compare with Eq. (\ref{eq:crosssection}))
\begin{equation}\label{eq:capture:DCSpartialwave}
\sigma_{l}=\dfrac{4\pi}{k^{2}}(2l+1)sin^{2}\eta_{l}. 
\end{equation}
In Eq. (\ref{eq:capture:DCSpartialwave}), the factor $sin^2(\eta_l)$ can now be identified with the probability $P_{l}$ for passing the centrifugal barrier \cite{clary90b}.  A quantum capture theory can thus be formulated by assuming that all partial waves up to a maximum value $l_{max}$ are fully reactive and higher partial waves are nonreactive \cite{clary85a}, i.e., they are reflected at the  barrier of the effective potential as in the classical capture picture. This means $P_{l} =1$ for $l\leq l_{max}$ and $P_{l} =0$ for $l>l_{max}$. With this condition, we get from Eqs. (\ref{eq:capture:DCSICS}) and (\ref{eq:capture:DCSpartialwave})
\begin{equation}\label{eq:capture:icsmax}
\sigma=\sum_{l=0}^{l_{max}}\sigma_{l}=\dfrac{4\pi}{k^{2}}(l_{max}+1)^{2}.
\end{equation}

Alternative formulations of quantum-capture theory can be derived by solving the scattering Schr\"odinger equation in a restricted region that contains the centrifugal barrier and the long-range part of the potential beyond \cite{clary90b}. The capture approximation is then made by assuming that there are no short-range forces that cause a back-reflection of reactive flux. 

Different approaches have been developed to account for the rotational motion of the neutral collision partner and a range of additional approximations have been introduced \cite{clary90b,armentrout04a}. These approximations are useful to reduce the computational cost of calculations and reveal the dominating physics behind the problem. These include adiabatic approximations  \cite{clary85b,clary95a,clary90b,clary87a,clary87a, troe87a,troe96a,dashevskaya03a}, the centrifugal sudden approximation (CSA) \cite{pack74a,maguire74a,clary85a}, the infinite-order sudden approximation (IOSA) \cite{tsien71a,pack72a,tsien73a,clary90b}, the average-dipole-orientation (ADO) approximation \cite{su73b,su75a,su73a}, the modified ADO theory taking into account the conservation of angular momentum  (AADO) \cite{su78a}, the adiabatic invariance method (AIM) \cite{bates81a,bates87a}, statistical approaches \cite{barker76b} and the perturbed-rotational state approximation \cite{kazuo78a,sakimoto80a,sakimoto81a,sakimoto82a}. 

In adiabatic capture theory, the Hamiltonian describing the problem is diagonalised in the basis of rotational wave functions of the neutral molecule \cite{clary85b}. This treatment yields a set of rotationally adiabatic potential curves based on which capture cross sections can be calculated \cite{clary85b,clary95a,clary90b,clary87a,clary87a}. The adiabatic capture (AC) approach is usually combined with other approximations to further simplify the problem \cite{clary84a}, e.g., the centrifugal sudden approximation (CSA) which ignores couplings between different $\Omega$ levels \cite{pack74a,maguire74a}, where $\Omega$ is the projection quantum number of the total angular momentum on the collision axis. This results in the so-called ACCSA method \cite{clary95a,clary87a}. It was shown \cite{ramillon94a} that ACCSA is formally equivalent to the treatment of ion-molecule capture within the statistical adiabatic channel model (SACM) developed by Troe \cite{troe85a,troe87a}.

Another treatment introduced is the infinite-order sudden approximation (IOSA) which assumes that cross sections can be calculated for fixed orientation angles $\theta$ of the colliding partners, and an average over all possible orientation angles and collision energies yields the rate constants. Combining IOSA and AC theory leads to a simple formulae for the cross section \cite{clary84a, clary90b}
\begin{equation}\label{eq:capture:ACIOSA}
\sigma = \dfrac{\pi}{2k}\int\limits_{0}^{\pi}d\theta sin(\theta)(J_{max}(\theta)+1)^{2}
\end{equation}
where $J_{max}(\theta)$ is the maximum total angular momentum for a given orientation angle $\theta$ for which capture is possible.


\subsection{Radiative processes}\label{ss:radiative}

In cold and dilute environments such as the interstellar medium or cold ion-atom mixtures (see Sec. \ref{ss:atomhybridtrap}), radiative chemical processes can play an important role. Besides photoexcitation and photoionization which are driven by the ambient radiation field, collisional radiative phenomena such as radiative charge transfer (RCT) and radiative association (RA) can become relevant.

\begin{figure}[!htb]
    \centering
    \resizebox{0.8\linewidth}{!}
    {\includegraphics{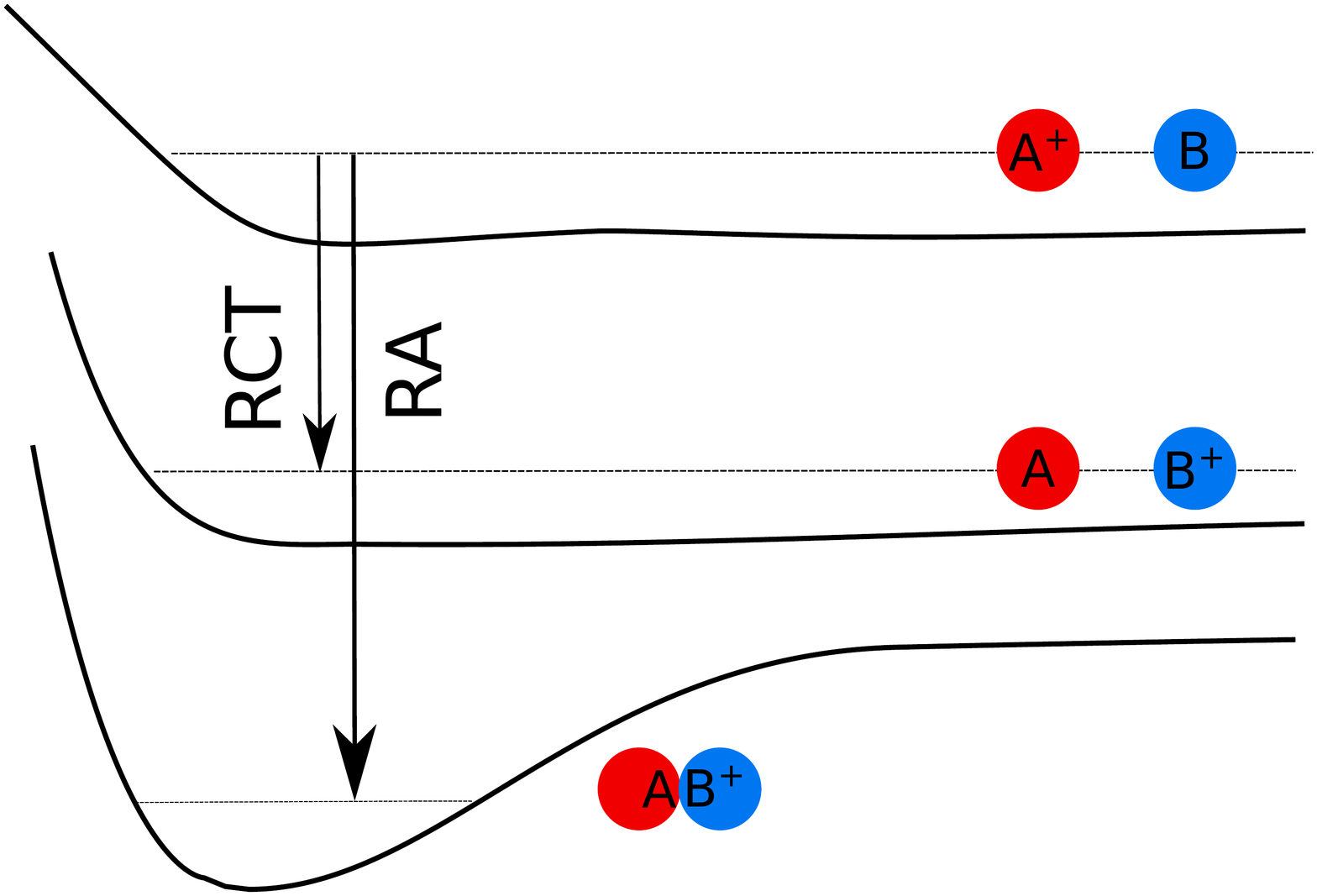}}
    \caption{Schematic representative of radiative charge transfer (RCT) and radiative association (RA) processes.
}
    \label{fig:RCTRA}
\end{figure}

RCT and RA occur in collisions between ions and neutrals when the collision complex emits a photon to populate levels associated with lower-lying molecular potential surfaces. Such processes can happen when the electron affinity of the ion (corresponding to the ionization energy of its neutral precursor) is larger than the ionization energy of the neutral collision partner. In this situation, there are necessarily always lower-lying electronic states of the collision system which connect to charge-transfer asymptotes, i.e., asymptotes where the charge has been exchanged between the collision partners. RCT occurs if the emission populates an unbound (scattering) state that connects to a charge-transfer asymptote. RA is a related process in which photon emission leads to the population of a lower-lying bound level, thus forming a molecular ion. A schematic presentation of RCT and RA processes is shown in Fig. ~\ref{fig:RCTRA}. 

The quantum-mechanical cross sections for RCT and RA can be calculated according to \cite{zygelman88a,hall13a,dasilva15a}:
\begin{eqnarray}
\sigma_{RCT}(E_{col})&=p&\frac{8\pi^2}{3c^3}\frac{1}{k^2} \sum_{J=0}^{\infty} \int_{0}^{E_{f}^{max}}\omega^3\left( J | \langle J-1, E_{f}|D(R)|E_{col},J\rangle |^{2}\right. \nonumber \\
&&\left.+(J+1) |\langle J+1, E_{f}|D(R)|E_{col},J\rangle |^{2}\right) dE_{f},
\label{csrct}
\end{eqnarray}
\begin{eqnarray}
\sigma_{RA}(E_{col})&=p&\frac{8\pi^2}{3c^3}\frac{1}{k^2} \sum_{J=0}^{\infty} \sum_{v=0}^{v_{max}}\left( \omega^3 J | \langle J-1,v|D(R)|E_{col},J\rangle |^{2}\right. \nonumber \\
&&\left.+ \omega^3  (J+1)| \langle J+1, v|D(R)|E_{col},J\rangle |^{2}\right), \label{csra} 
\end{eqnarray}
Here, $p$ is the statistical weight of the entrance channel of the collision, $k$ is the collisional wavenumber, $E_f$ is the relative energy of the products, $\omega$ is the frequency of the emitted photon, $J$ is the total angular momentum quantum number, $D(R)$ is the electronic transition dipole moment between the initial and final state and $v$ is the vibrational wavenumber of the product molecular ion. The quantities in these equations are usually derived from $ab$-$initio$ or similar calculations of the potentials. Note that the two equations above differ by the final wave functions, which is either the one for a continuum state $\langle J-1, E_{f}|$, or for a bound state $ \langle J-1,v|$.

\section{Experimental methods}\label{se:methods}


\subsection{Ions in flows and beams}\label{ss:beam}

\begin{figure}[!htb]
    \centering
    \resizebox{0.8\linewidth}{!}
    {\includegraphics{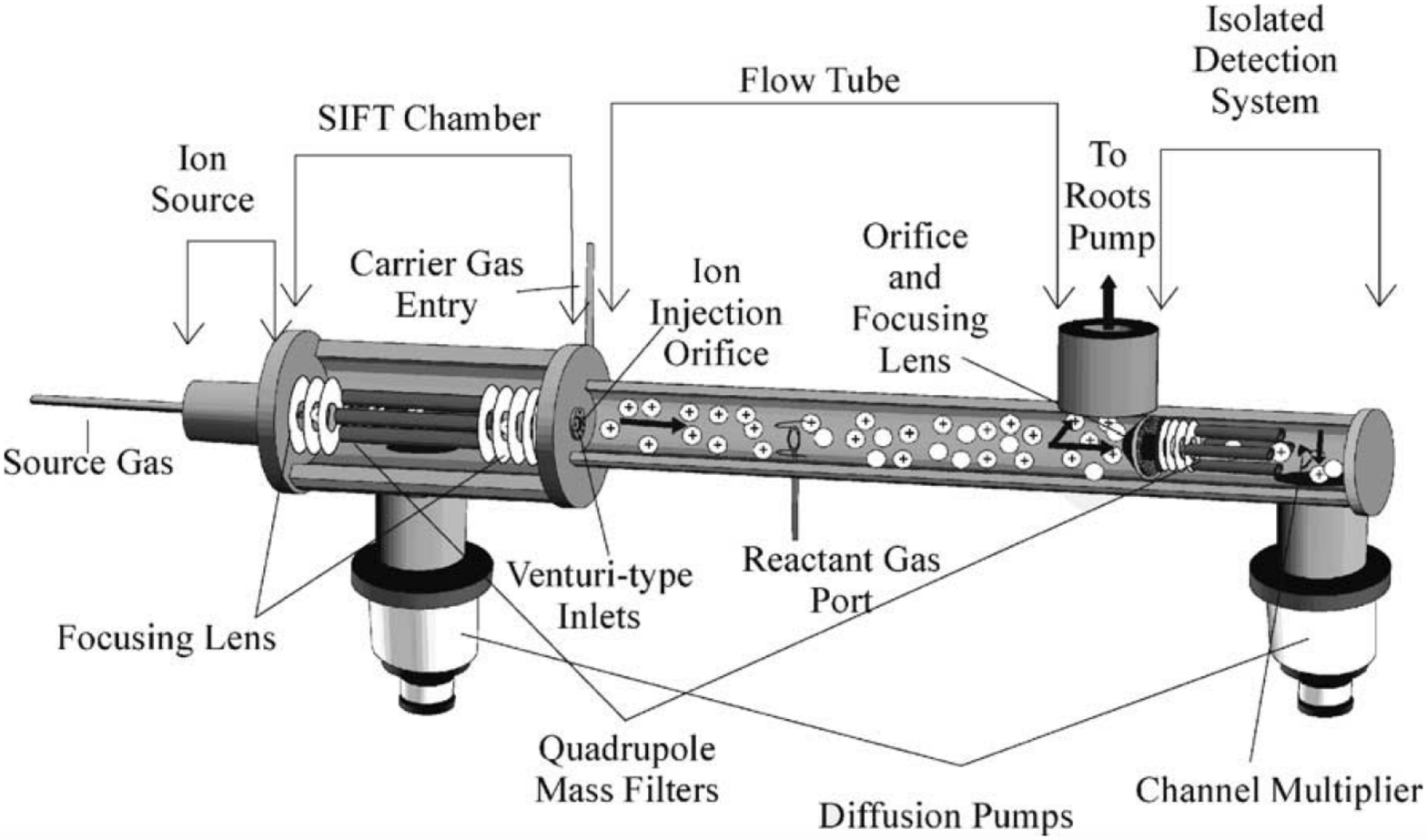}}
    \caption{Schematic diagram of a SIFT apparatus. Ions are focused from an ion source into a quadrupole mass filter for mass selection. The mass-selected ions are injected into a flow tube to be entrained in a flow of He carrier gas to which reactant gases are added. Reactant and product ions are sampled downstream by a mass filter and ion detection system. Reprinted from Int. J. Mass Spectrom., Vol 243, Douglas M. Jackson, Nathan J. Stibrich, Nigel G. Adams and Lucia M. Babcock, A selected ion flow tube study of the reactions of a sequence of ions with amines, 115--120, Copyright (2005), with permission from Elsevier.
}
    \label{fig:sift}
\end{figure}

Over the years, considerable efforts have been devoted to the development of methods for studies of ion-molecule reactions at low temperatures. Early experiments made use of the flowing afterglow (FA) method \cite{ferguson69a} and its further development such as selected ion flow tubes (SIFT) \cite{adams76a}. The SIFT technique eliminated many sources of interference in ion-molecule reaction experiments such as the presence of electrons, energetic photons, neutral fragments and excited species. This was accomplished by using a mass filter to select a single ionic species from a remote ion source and injecting it, at low energy, into a flow of neutral gas. The ions thermalize in the flow and react with gases added downstream of the injection point. Reactions are studied using mass-spectrometric detection of reactants and products. A schematic diagram of a SIFT apparatus is displayed in Fig. \ref{fig:sift} \cite{Jackson05a}. Early SIFT experiments covered the temperature range down to $\approx$ 80 \si{\kelvin} by cooling the main flow tube \cite{adams76a,adams79a,ferguson84a,clary85b,
henchman89a}. Temperatures as low as 18 \si{\kelvin} have been reached by in the SIFT technique by using liquid He as a coolant \cite{bohringer86a}.

Another major development was the cin\'{e}tique de r\'{e}actions en ecoulement supersonique uniforme (CRESU) technique \cite{rowe84a}, 
described in detail in chapter XX. The crucial feature of this technique is the isoentropic expansion of a gas, primarily composed of a nonreactive species 
into which the reactants are seeded.
This method was employed to study the kinetics of ion-molecule reactions at temperatures down to $\approx$ 8 \si{\kelvin} achieved by precooling the gas prior to its expansion \cite{rowe85a}.
CRESU experiments were originally performed in a configuration similar to that used in SIFT experiments: Ions were created by irradiating the gas just outside the Laval nozzle using an electron beam. They, and any product ions from reactions with the neutral gas, could be detected further downstream with a mass spectrometer. The observation of variations in the ion signals, taking into account the composition of the flowing gas and the distance of the sampling point from the nozzle, were used to determine rate coefficients. In more recent experiments, the CRESU apparatus was modified to allow the selection of the ionic species \cite{rebrion89a}, and the generation of ions by resonance-enhanced multiphoton ionization \cite{canosa11a,schlappi15a}. 

\begin{figure}[!htb]
    \centering
    \resizebox{0.8\linewidth}{!}
    {\includegraphics{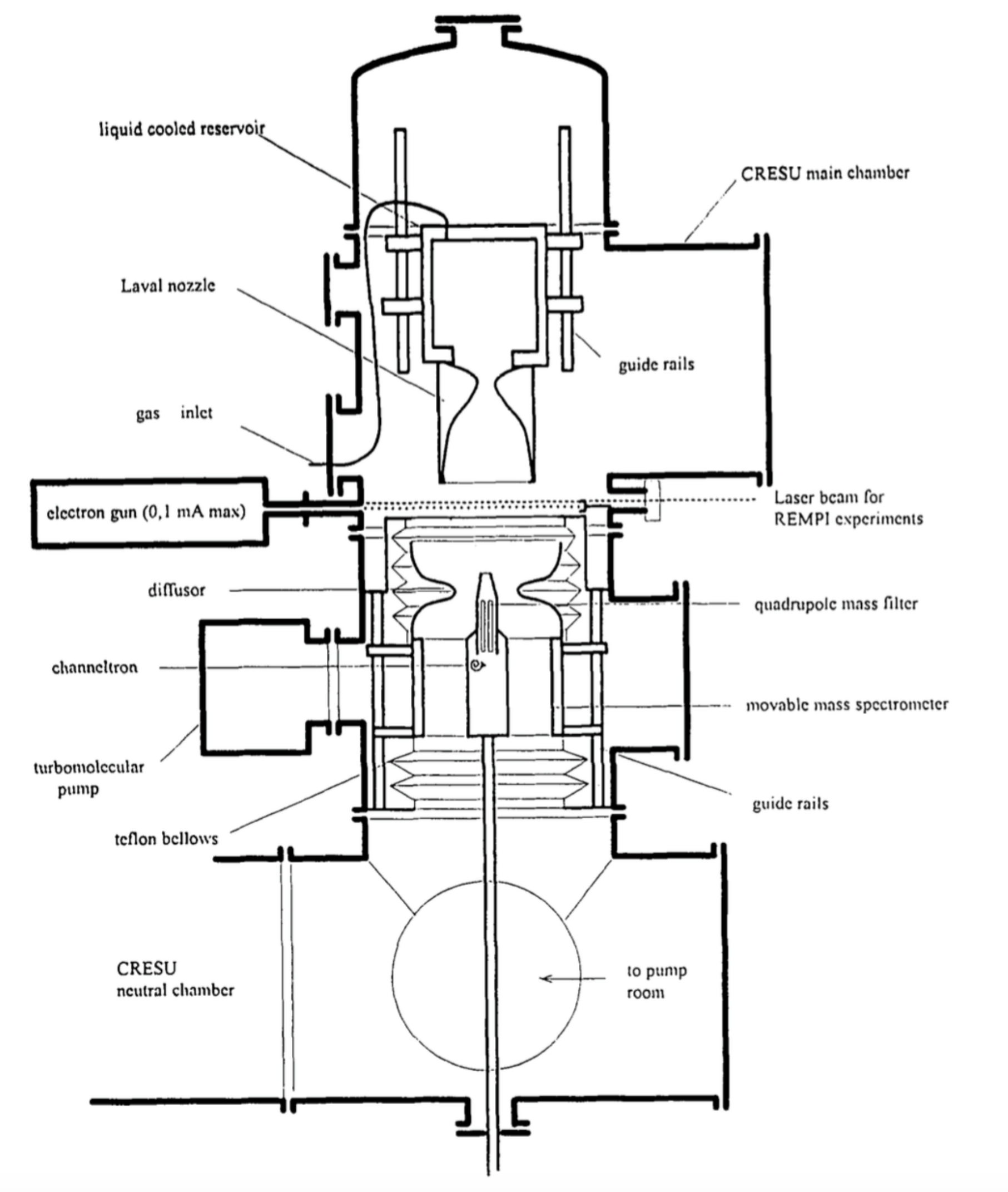}}
    \caption{Sketch of a CRESU (Cin\'{e}tique de R\'{e}action en Ecoulement Supersonique Uniforme) apparatus devoted to the study of ion-molecule reactions. Figure adapted from Int. J. Mass Spectrom. Ion Proc., Vol 149/150, B. R. Rowe and A. Canosa and V. \mbox{Le Page}, FALP and CRESU studies of ionic reactions, 573, Copyright (1995), with permission from Elsevier. 
}
    \label{fig:cresu}
\end{figure}

Cold ion-molecule reactions have also been studied in pulsed free jet expansions in which short pulses of gas are expanded into high vacuum through a pinhole-type nozzle \cite{smith98a}. Ions are usually generated in the throat of the nozzle by methods such as electron-impact ionization or photoionization. The pulsed operation mode allows the generation of higher pressure differentials between the gas reservoir and the expansion chamber, thus improving cooling and enabling gas temperatures of a few Kelvin in the moving frame of the jet. One disadvantages of this approach is the rapidly decreasing gas density with distance from the nozzle, which implies that collisions can only be studied within a few nozzle diameters from the orifice. As a consequence, the molecules are usually not in thermal equilibrium along the extension of the jet and temperature gradients have to be accounted for in the analysis of the data. 


\subsection{Ion trapping}\label{ss:trap}

An alternative method for studies of cold ion-molecule reactions is the ion-trap technique. Compared to the flow and beam methods discussed in the previous section, despite the low number of particles achievable, ion traps have the advantage of offering longer observation times, of achieving lower temperatures and of providing additional facilities for controlling the ions \cite{gerlich92a, gerlich95a, willitsch12a, willitsch16a, major05a,march05a}.

Radiofrequency (RF) traps \cite{gerlich92a, drewsen00a, major05a, march05a} have been established as the major type of instruments used for the storage of ions at low temperatures. These devices confine ions in inhomogeneous electric fields $\Vec{E}$
\begin{equation}\label{eq:iontrap:Efield}
\vec{E}(\vec{r},t)= \vec{E_{0}}(\vec{r})cos(\Omega t),
\end{equation} 
with frequencies $\Omega$ in the \si{\mega\hertz} regime. Here, $\Vec{E}_0$ is the field amplitude vector and $t$ is the time. The oscillating fields create a dynamic trap in which the ions oscillate around the trap minimum. To stably confine ions in the trap, the stability parameter $\eta$, defined by
\begin{equation}\label{eq:iontrap:stabilitypaprameter}
\eta=\dfrac{2Q\vert\nabla \vec{E_{0}}\vert}{m\Omega ^{2}},
\end{equation}
must be smaller than $\approx$ 0.3 (as shown in numerical simulations) in the volume sampled by the ion motion \cite{gerlich92a, major05a,wester09a}. In Eq. (\ref{eq:iontrap:stabilitypaprameter}), $Q$ is the charge of the ion and $m$ is its mass. In this regime, the motion of the ions in the RF trap can be described by a slow thermal ("secular") motion superimposed by a fast "micromotion" caused by the rapid oscillation of the RF potential \cite{major05a}. Thus, the time-averaged kinetic energy stored in the micromotion equals a pseudopotential and can be formulated as
\begin{equation}\label{eq:iontrap:pseudopotential}
V^{\ast}(\vec{r})=\dfrac{Q^{2}\Vec{E}_{0}(\Vec{r})^{2}}{4m\Omega^{2}}
\end{equation}
which governs the secular motion of the ions in the trap \cite{gerlich92a, major05a, march05a}. 

\begin{figure}[!htb]
    \centering
    \resizebox{0.8\linewidth}{!}
    {\includegraphics{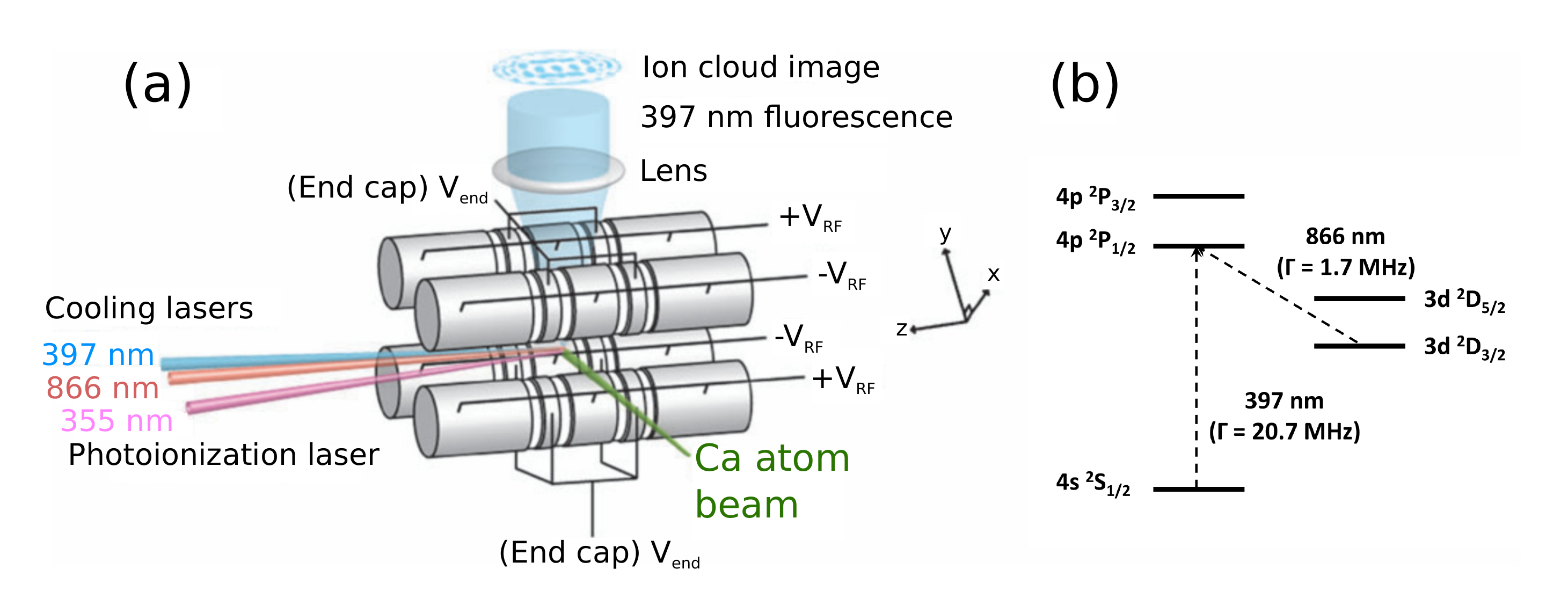}}
    \caption{(a) Schematic of a linear radiofrequency quadrupole trap used for the storage and laser cooling of Ca$^+$ ions. Static ($V_{end}$) and radiofrequency ($V_{RF}$) voltages applied to the electrodes are used to generate a dynamic trap for the ions. Laser beams at 355~nm and 397~nm as well as 866~nm serve to generate and cool the ions. The fluorescence generated during laser cooling is imaged onto a camera using a microscope. (b) Diagram of the energy levels relevant for the Doppler laser cooling of $^{40}$Ca$^{+}$. The main cooling transition is $(4s)~^2S_{1/2} \rightarrow (4p)~^2P_{1/2}$ at 397~nm. Another laser at 866~nm is required to repump population trapped in the metastable $(3d)~^2D_{3/2}$ level in order to close the laser-cooling cycle. Reproduced from Ref. \cite{willitsch08b} with permission from the PCCP Owner Societies.
}
    \label{fig:paultrap}
\end{figure}

One of the most common types of RF traps used to date is the linear Paul (or quadrupole) trap \cite{prestage89a}. This device typically consists of four segmented cylindrical electrodes arranged in a quadrupolar configuration, see Fig. \ref{fig:paultrap} (a). The polarity of the RF voltages $V_{RF}=V_0cos(\Omega t)$ with amplitude $V_0$ alternates between adjacent electrodes. This configuration creates an oscillating quadrupolar potential in the direction perpendicular to the logitudinal trap axis which dynamically confines the ions. Additionally, the ions are trapped along the longitudinal direction by static electric fields applied to the outermost segments (the end caps) as shown in Fig. \ref{fig:paultrap} (a), thus creating a three-dimensional trap. 

\begin{figure}[!htb]
    \centering
    \resizebox{0.6\linewidth}{!}
    {\includegraphics{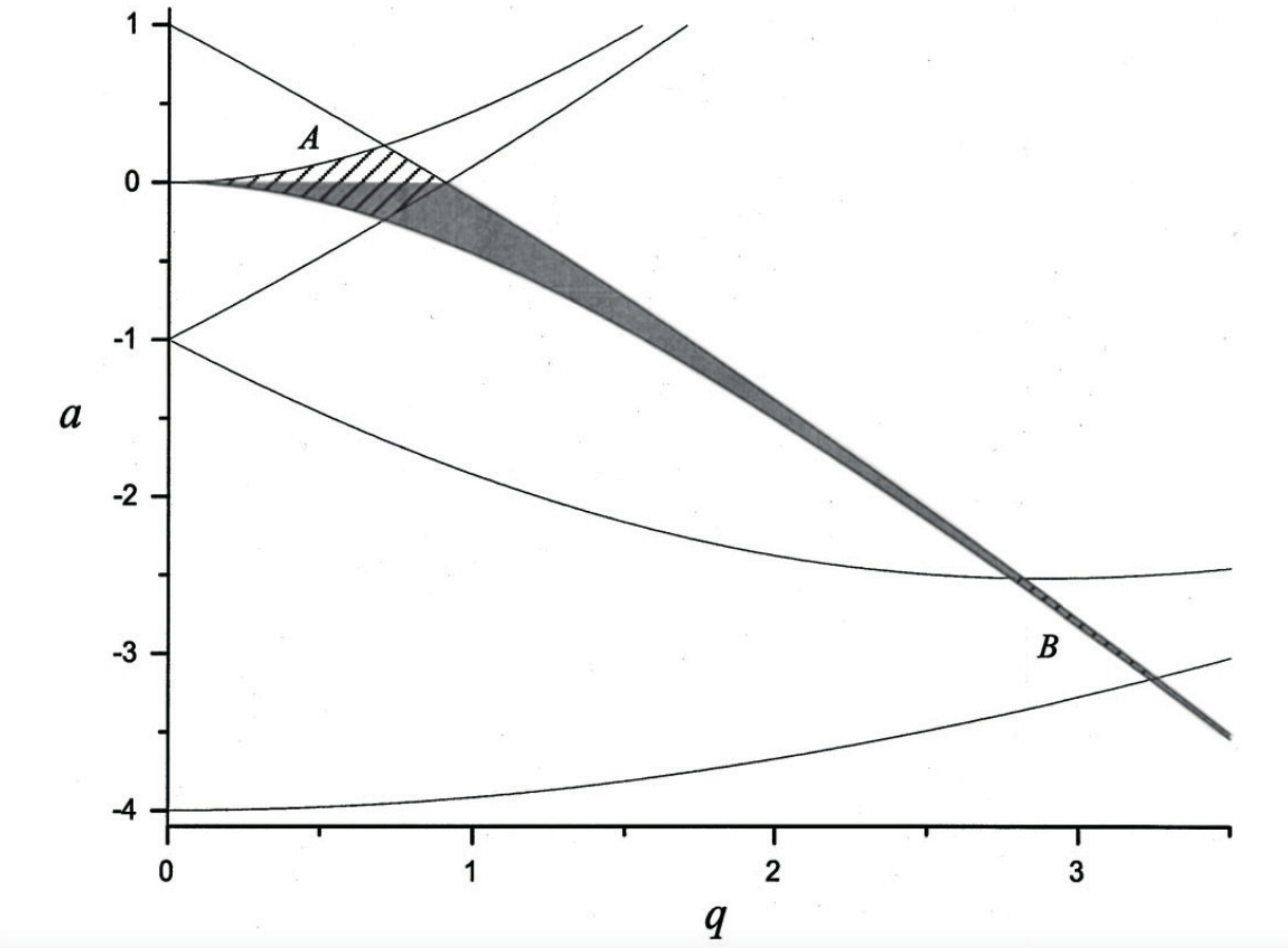}}
    \caption{Stability diagram for a linear Paul trap. The overlap between grey-shaded area and the hatched areas A and B (stable region for quadrupole  mass
filter) indicates a "region of stability", i.e., combinations of Mathieu parameters $a$ ($a_{u}$) and $q$ ($q_{u}$) for which an ion is stably confined in the trap. The Mathieu parameters are defined in Eqs. (\ref{eq:Mathieuparameter1}) and (\ref{eq:Mathieuparameter2}) of the text. Reprinted with permission from M. Drewsen and A. Br\o ner, Phys. Rev. A 62, 045401 (2000). Copyright (2000) by the American Physical Society.
}
    \label{fig:stabilitydiagram}
\end{figure}

The motion of a single ion in a linear quadrupole trap can be described by a set of Mathieu equations \cite{drewsen00a, major05a, march05a}
\begin{equation}\label{eq:Mathieu} 
\dfrac{d^{2}u}{d\tau^{2}}+[a_{u}+2q_{u}(2\tau)]u=0,
\end{equation}
where $u$ is one of the cartesian coordinates $x$, $y$ or $z$ and $\tau = (1/2)\Omega t$. 
Stable confinement of the ion can be achieved by a proper choice of the trapping conditions expressed as the Mathieu parameters
\begin{equation}\label{eq:Mathieuparameter1}
 a_{x}=a_{y}=-\dfrac{1}{2}a_{z}=-k\dfrac{4QV_{end}}{m\Omega^{2}z_{0}^{2}},
\end{equation}
\begin{equation}\label{eq:Mathieuparameter2}
 q_{x}=-q_{y}=\dfrac{2QV_{0}}{m\Omega^{2}r_{0}^{2}}, q_{z}=0.
\end{equation}
where $2z_{0}$ is the distance between two end caps, $r_{0}$ is the radius of the circle inscribed by the quadrupole electrodes, and $V_{end}$ is the static voltage on the end caps. $k$ is a factor related to the geometry of the ion trap. The Mathieu equations can be solved analytically to obtain the trajectory of the ion \cite{major05a}. It can be shown that the ion will remain stored in the RF trap if the values of the Mathieu parameters $a_{u}$ and $q_{u}$ fall into a "region of stability"
indicated as the grey area in the diagram shown in Fig. \ref{fig:stabilitydiagram} \cite{drewsen00a}.

\begin{figure}[!htb]
    \centering
    \resizebox{0.8\linewidth}{!}
    {\includegraphics{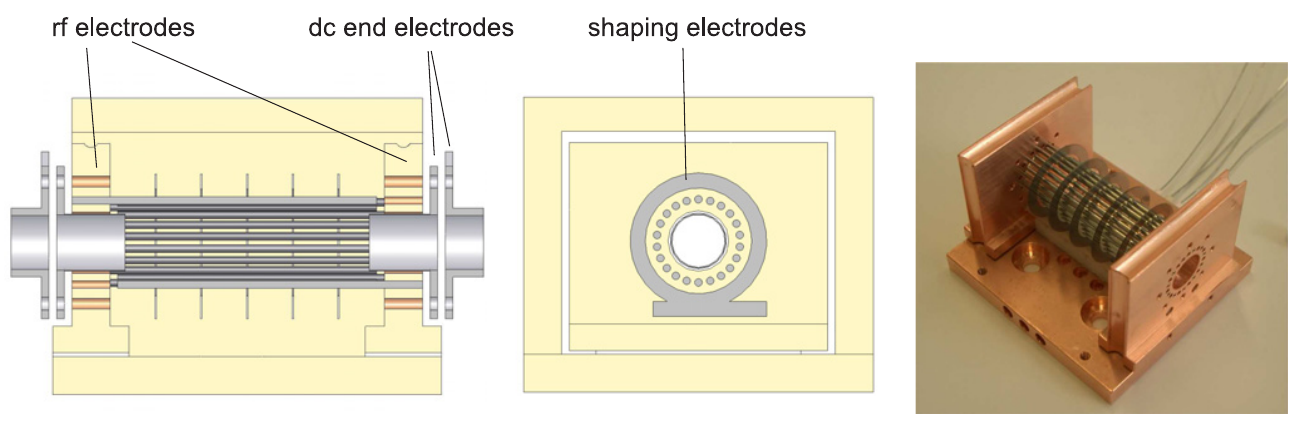}}
    \caption{A linear 22-pole radiofrequency ion trap used for the cooling of ions with a cryogenic buffer gas. The 22 rf electrodes, mounted alternately into the two
side plates and two pairs of dc end electrodes provide ion confinement. The shaping electrodes shown in the middle figure are used to provide additional axial trapping potential and to improve the mass resolution for later on detection \cite{Midosch08a}. From Ref. \cite{wester09a}. Copyright (2009) by IOP Publishing. Reproduced by permission of IOP Publishing. All rights reserved. 
    }
    \label{fig:22pole}
\end{figure}

Apart from quadrupole traps, linear higher-order-multipole RF traps \cite{gerlich09a, wester09a} have recently found a wide range of applications, in particular in the context of buffer-gas cooling (see the following section). These traps consist of $2n$ cylindrical electrodes, with integer $n>0$, evenly distributed around a circle. An ideal linear multipole with oscillating voltages $V_{RF}=V_0cos(\Omega t)$ applied with opposite polarity to adjacent rods generates an electric potential of the form 
\begin{equation}
V(r,\phi,t)=V_0cos(n\phi)(r/r_0)^nsin(\Omega t)
\label{eq:multipole}
\end{equation}
in the plane perpendicular to the multipole axis. Eq. (\ref{eq:multipole}) is given in cylindrical coordinates with $r$ and $\phi$ the radial and angular coordinate, respectively. $r_0$ is the inscribed radius of the multipole. This electrode configuration creates a pseudopotential of the form
\begin{equation}
V^*(r)=\frac{Q^2n^2V_0^2}{4m\Omega^2r_0^2}\left(\frac{r}{r_0}\right)^{2n-2}.
\label{eq:multipolepseudopot}
\end{equation}
For $n=2$, one recovers the harmonic pseudopotential of a linear quadrupole trap. With increasing multipole order $n$, the pseudopotential becomes increasingly shallower close to the center of the trap and steeper towards the electrodes, i.e., it becomes increasingly box-shaped. Thus, the effective volume of the trap is increased. A convenient experimental implementation of a higher-order multipole trap is a linear 22-pole, see  Fig. \ref{fig:22pole}, originally introduced by Gerlich \cite{gerlich95a}.


\subsection{Buffer gas cooling}\label{ss:buffergas}

An important method for the generation of cold trapped ions for subsequent studies of chemical reactions is buffer-gas cooling \cite{gerlich09a,hutzler12a,quemener12a}. Cryogenic buffer gases act as a coolant for the ions by thermalization of their translational and internal degrees of freedom through elastic and inelastic collisions. There are several criteria for the successful application of buffer gas cooling: First, the elastic collision cross sections between the buffer gas and the ions should be large such that buffer gas cooling is efficient. Second, inelastic and reactive collision cross sections should be small and the buffer gas should not react with the ions. Third, the mass of the buffer gas should be smaller than the mass of the ions \cite{dehmelt68a}, although this restriction can be relaxed under certain conditions \cite{hoeltkemeier16a}. This last criterion is linked to the problem of RF heating, i.e., the coupling of micromotion energy into the thermal motion of the ions by collisions with the buffer gas \cite{dehmelt68a, zipkes11a, cetina12a, chen14a, chen14a, hoeltkemeier16a}. The RF heating rates decrease with increasing multipole order $n$ of the trap, because of the shallower pseudopotential and therefore reduced micromotion amplitudes across the trapping volume of the ions \cite{gerlich92a, wester09a}. For these reasons, buffer-gas cooling experiments are mostly performed in high-order multipole traps such as 22-poles. 

\begin{figure}[!htb]
    \centering
    \resizebox{0.8\linewidth}{!}
    {\includegraphics{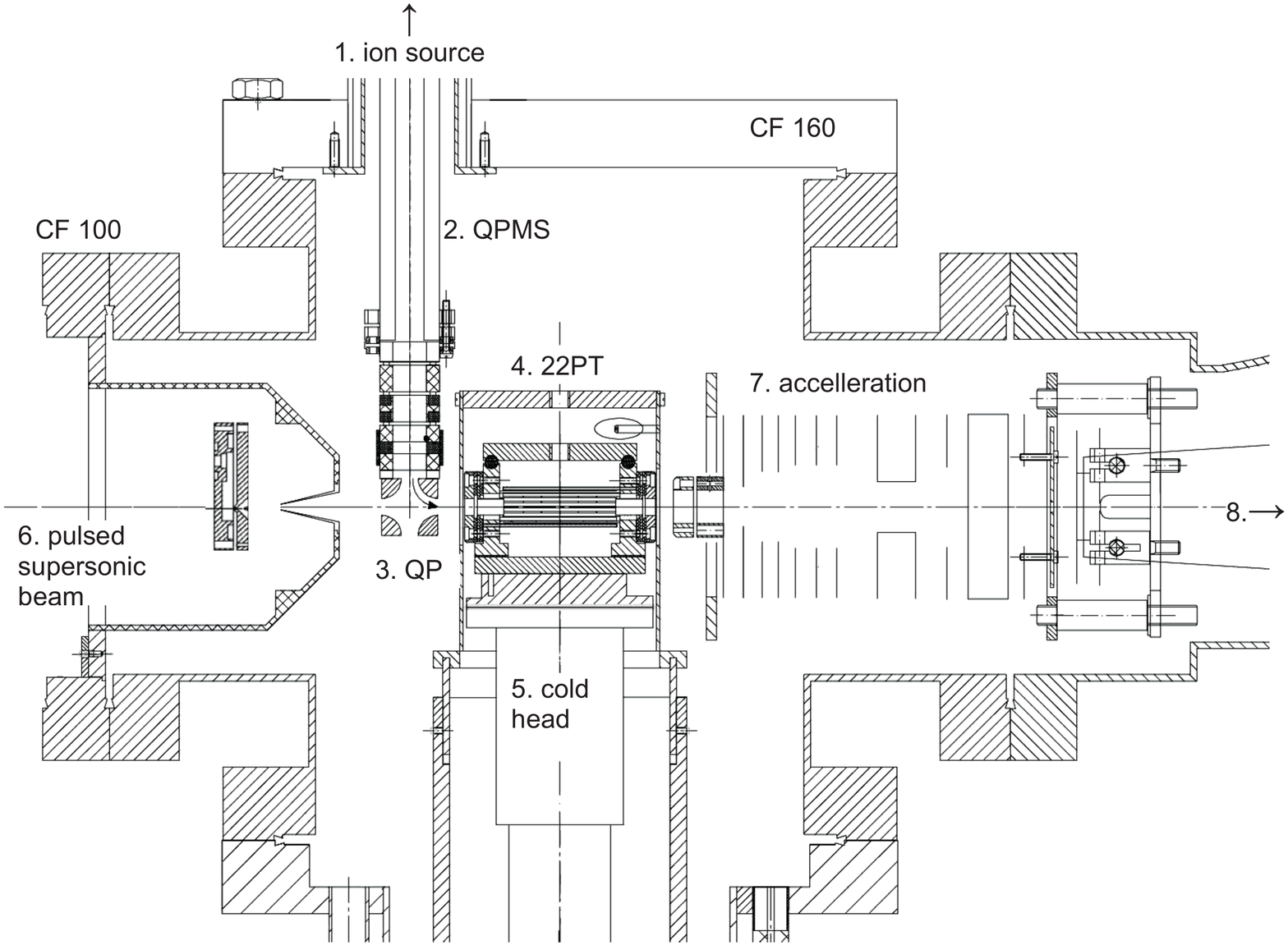}}
    \caption{Construction details of a temperature-variable 22-pole RF ion trap coupled to a pulsed supersonic molecular beam. Primary ions are produced in a storage ion source (1, not shown), mass selected in a quadrupole mass analyser (2), deflected in a quadrupole bender (3) and injected into the 22-pole trap (4) which is mounted on the two stages of a cold head (5, variable from 4 K to room temperature). In the trap, the ions are confined by the RF field and static voltages applied to the entrance and exit electrodes. Buffer gas, usually He, is used for thermalizing the ions. From the left, a skimmed molecular beam (6) traverses the trap without colliding with the cold surfaces to provide neutral molecules for reactions with the ions. For detection, the stored ions are extracted to the right, accelerated (7), mass selected in a magnetic field (8), and counted via an ion detector. Reprinted from Ref. \cite{gerlich06a}. Copyright (2006) by the Royal Swedish Academy of Sciences. Reproduced by permission of IOP Publishing. All rights reserved.
}
    \label{fig:buffergas}
\end{figure}

The most widely used buffer gas is cryogenic He which is used because of its high vapor pressure at cryogenic temperatures and its small mass. In typical experimental setups as shown in Fig. \ref{fig:buffergas} \cite{gerlich95a,gerlich06a, wester09a}, the trap and an enclosing cryogenic shield are thermally connected to a suitable cryogenic system such as a closed-cycle refrigerator. The buffer gas is injected into the trapping region to thermalize with the cold surfaces and in turn to cool trapped ions by collisions. Depending on the cryogenic source, nominal temperatures as low as a few \si{\kelvin} can been reached, although the effective ion temperature will usually be higher than the bath temperature because of RF heating \cite{wester09a}.


\subsection{Laser and sympathetic cooling}\label{ss:lasercooling}

Laser cooling \cite{metcalf99a,foot05a,eschner03a} is an alternative method for the preparation of cold, trapped atomic ions (Note that it is also widely used for neutral atoms). This approach is based on the momentum transferred to atoms during the absorption of photons from a laser beam. In these experiments, the frequency of the laser is slightly tuned to the red of an atomic resonance. Atoms with a velocity component opposite to the propagation direction of the laser beam experience a blue shift of the laser frequency due to the Doppler effect. This brings the laser frequency closer to resonance in the frame of the atom and increases the probability for the absorption of photons. The momentum transfer from a photon absorbed in this configuration leads to a slight deceleration of the atom. The excited state of the atom subsequently decays by spontaneous emission of a photon into a random direction. In such a way a "closed optical cycle" can be formed. Averaged over many repetitions, the momenta transferred by the emitted photons average out. The atom will thus receive only a net momentum transferred by the photons absorbed from the laser beam which slows it down. For large ensembles of trapped ions only one laser beam is usually sufficient for cooling the ions' motions in all three spatial directions because Coulomb coupling leads to a thermalization of all translational degrees of freedom. Laser-cooling experiments are usually performed in linear quadrupole traps which provide a stiff, harmonic trapping potential for the ions. 

The lowest translational temperature which can be reached with Doppler laser cooling in the absence of other cooling mechanisms is given by
\begin{equation}\label{eq:lasercooling:Dopplerlimit}
T_{D}=\dfrac{\hbar\Gamma}{2k_{B}},
\end{equation} 
where $\Gamma$ is the natural linewidth of the relevant atomic transition. Laser cooling is usually performed on systems with a simple energy-level structure, such as singly charged alkaline-earth ions, which facilitates the implementation of closed optical cycles. For instance, for the Ca$^{+}$ ion the $(4s)~^2S_{1/2} \rightarrow (4p)~^2P_{1/2}$ transition is frequently used for laser cooling, see Fig. \ref{fig:paultrap} (b). The corresponding natural line width is 20 \si{\mega\hertz} and the Doppler limit is $T_{D}$ = 0.5 \si{\milli\kelvin}. 

At sufficiently low ion kinetic energies, the cloud of trapped ions undergoes a phase transition to an ordered state, a Coulomb crystal \cite{molhave00a,drewsen03a,willitsch08b,willitsch12a,heazlewood15a}. This process can be quantified by the plasma-coupling parameter $\Gamma$ which is defined as the ratio of the Coulomb energy $E_C$ to the kinetic energy $E_K$ of the ions: 
\begin{equation}\label{eq:plasma-coupling}
\Gamma=\dfrac{E_{C}}{E_{K}}=\dfrac{Q^{2}}{4\pi\varepsilon_{0}a_{WS}k_{B}T}.
\end{equation}
In Eq. (\ref{eq:plasma-coupling}), $a_{WS}=(3/(4\pi\rho))^{1/3}$ is the Wigner-Seitz radius which denotes the average distance between the ions, where $\rho$ is the ion density. Theoretical calculations have predicted that Coulomb crystallization occurs if the plasma-coupling parameter exceeds a value of $\Gamma \simeq 150$ \cite{pollock73a, farouki93a}.

\begin{figure}[!htb]
    \centering
    \resizebox{0.8\linewidth}{!}
    {\includegraphics{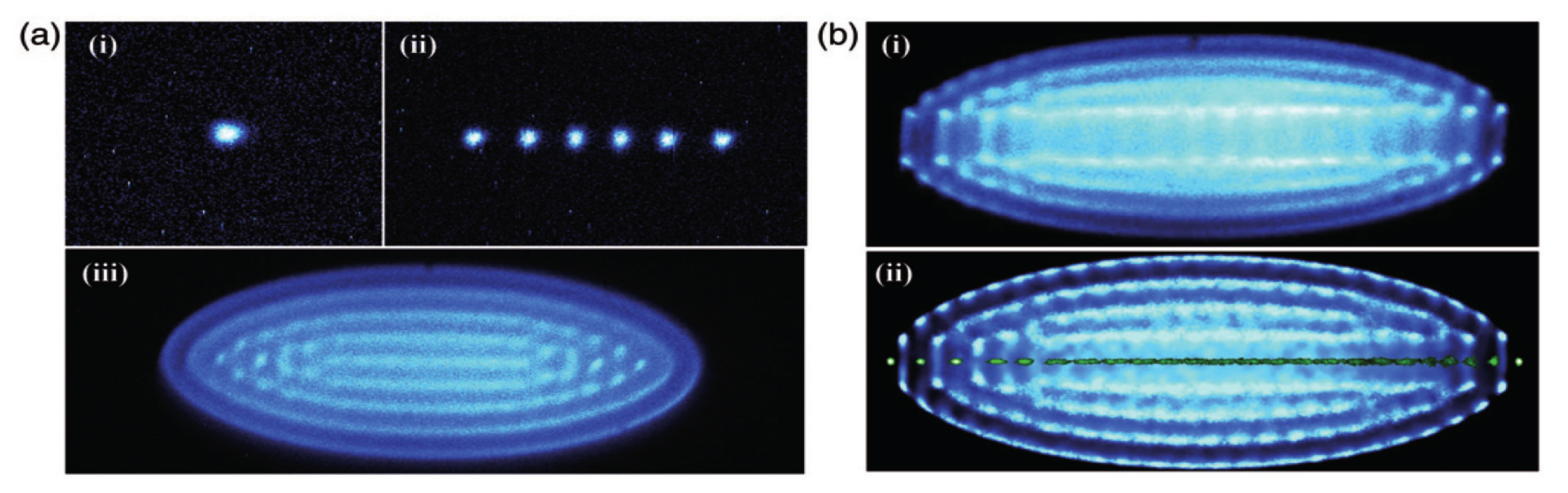}}
    \caption{False-colour fluorescence images of Coulomb crystals of laser-cooled Ca$^{+}$ ions in a linear Paul trap. (a) (i) A single laser-cooled ion, (ii) a string of six ions, (iii) a large spheroidal Coulomb crystal (not to scale with the images shown in (i) and (ii)). (b) (i) Bicomponent Coulomb crystal consisting of 925 laser-cooled Ca$^{+}$ ions and 24 sympathetically cooled N$_{2}^{+}$ ions. The molecular ions form a non-fluorescing region in the center of the crystal. (ii) Molecular-dynamics simulation of the image in (i). The distribution of molecular ions in the crystal has been made visible in green in the simulation. Figure adapted from Ref. \cite{willitsch12a}: Coulomb-crystallised molecular ions in traps: methods, applications, prospects, Stefan Willitsch, International Reviews in Physical Chemistry, published on 27 Mar 2012. Reprinted by permission of Taylor \& Francis.}
    \label{fig:cc}
\end{figure}

Coulomb crystals have several intriguing properties. First, the ions are tightly localized so that single particles can be observed and manipulated, see Fig. \ref{fig:cc}. In this figure, each dot shows a single ion that is visible due to the light scattered during laser cooling. This provides an attractive platform for reaction studies on isolated particles. Second, the ions are cold. The \si{\milli\kelvin} regime can readily be reached with Doppler laser cooling. In small ensembles consisting of only a few ions, even the $\mu$K regime can be accessed with advanced techniques such as resolved-sideband cooling or EIT (electromagnetically induced transparency) cooling \cite{haeffner08a, lechner16b}. Third, the ions can be stored for time scales of minutes and hours, enabling long experiment times on a single sample of particles.

Compared to atomic ions, molecular ions can, in most cases, not be laser cooled due to the lack of the closed optical transitions which can be cycled repeatedly, with possible exceptions \cite{nguyen11a}. However, cold molecular ions can be prepared using co-trapped, laser-cooled atomic ions as coolant (sympathetic cooling \cite{wineland80a,wineland86a,molhave00a}). Via elastic collisions between the trapped particles, kinetic energy is transferred from the hot molecular ions to the cold atomic ions from where it is ultimately removed by laser cooling. As a result, the molecular and atomic ions are simultaneously Coulomb crystallized in the ion trap forming bi-component Coulomb crystals. Since the first realization of bicomponent Coulomb crystals \cite{molhave00a}, sympathetic cooling has become the routine method for the generation of cold molecular ions in traps. The technique is widely applicable to a variety of species ranging from atomic ions \cite{wineland86a,bowe99a,hornekaer01a}, diatomic and small polyatomic molecular ions \cite{molhave00a,roth06b,roth06c,tong10a} to large organic and biomolecular ions \cite{ostendorf06a,hojbjerre08a,offenberg08a}. Fluorescence images of single- and bicomponent crystal structures are shown in Fig. \ref{fig:cc}. 

In general, the efficiency of sympathetic cooling is optimal if the two ion species have similar mass-to-charge ratios $m/z$, in which case the secular temperature of the molecular ions is close to the one of the atomic species \cite{tong10a}. However, even in extreme cases such as the sympathetic cooling of multiply-charged cytochrome c ions (mass $\simeq$ 12390~u, charge +17~$e$) with $^{138}$Ba$^{+}$, secular temperatures as low as 750 \si{\milli\kelvin} have been obtained \cite{offenberg08a}.


\subsection{State-preparation of neutral molecules and  ions}\label{ss:state-preparation}

In the context of cold chemistry, it is not only important to control the translational motion of the reaction partners, but also their internal quantum states. Hence, many cold-chemistry experiments aim to combine translational cooling with the cooling or even specific preparation of the internal states of the neutral and ionic reactants.

For decades, the working horse method to generate internally cold samples of neutral molecules has been their adiabatic cooling in supersonic gas expansions leading to the formation of molecular beams \cite{scoles88a}. With state-of-the-art pulsed gas nozzles \cite{even15a}, rotational temperatures on the order of 1~K can be achieved. The supersonic expansion also leads to a compression of the velocity distribution and therefore translational cooling in the moving frame of the gas pulse, but the molecules remain fast in the laboratory frame.

In recent years, various methods have been devised to slow down pulsed molecular beams using external electric, magnetic and radiation fields to create translationally cold molecules in the laboratory frame \cite{hogan11a,narevicius12a,meerakker08a,carr09a,meerakker12a,fulton04a}. One of the most important techniques in this domain is Stark deceleration which is based on the slowing of molecules in time-dependent inhomogeneous electric fields \cite{meerakker08a, meerakker12a}. Because the interaction of the molecules with the field (the Stark effect) is dependent on their internal state, Stark deceleration may produce molecular samples with a high degree of state purity under certain conditions, see Refs. \cite{meerakker12a, fitch12a} for details. A conceptually similar technique is Zeeman deceleration in which molecular beams containing paramagnetic species are decelerated using time-dependent inhomogeneous magnetic fields \cite{vanhaecke07a,Narevicius07a,hogan11a,narevicius12a}. 

Similarly, a range of different methods has been developed for the cooling and preparation of the internal states of molecular ions. Cooling of trapped ions with a  buffer gas, as described in Sec. \ref{ss:buffergas}, also affects the internal degrees of freedom so that the rotational and vibrational motions usually thermalize at temperatures close to the one of the cryogenic gas. On the other hand, sympathetic cooling of molecular ions with laser-cooled atomic ions, as discussed in Sec. \ref{ss:lasercooling}, chiefly relies on the interaction between electrostatic monopoles which does not couple to the external rotational and vibrational degrees of motion of the ions within the trap. As a consequence, sympathetically cooled molecular ions are usually translationally cold, but internally warm in equilibrium with the ambient thermal radiation field \cite{bertelsen06a}. 

\begin{figure}[!htb]
    \centering
    \resizebox{0.6\linewidth}{!}
    {\includegraphics{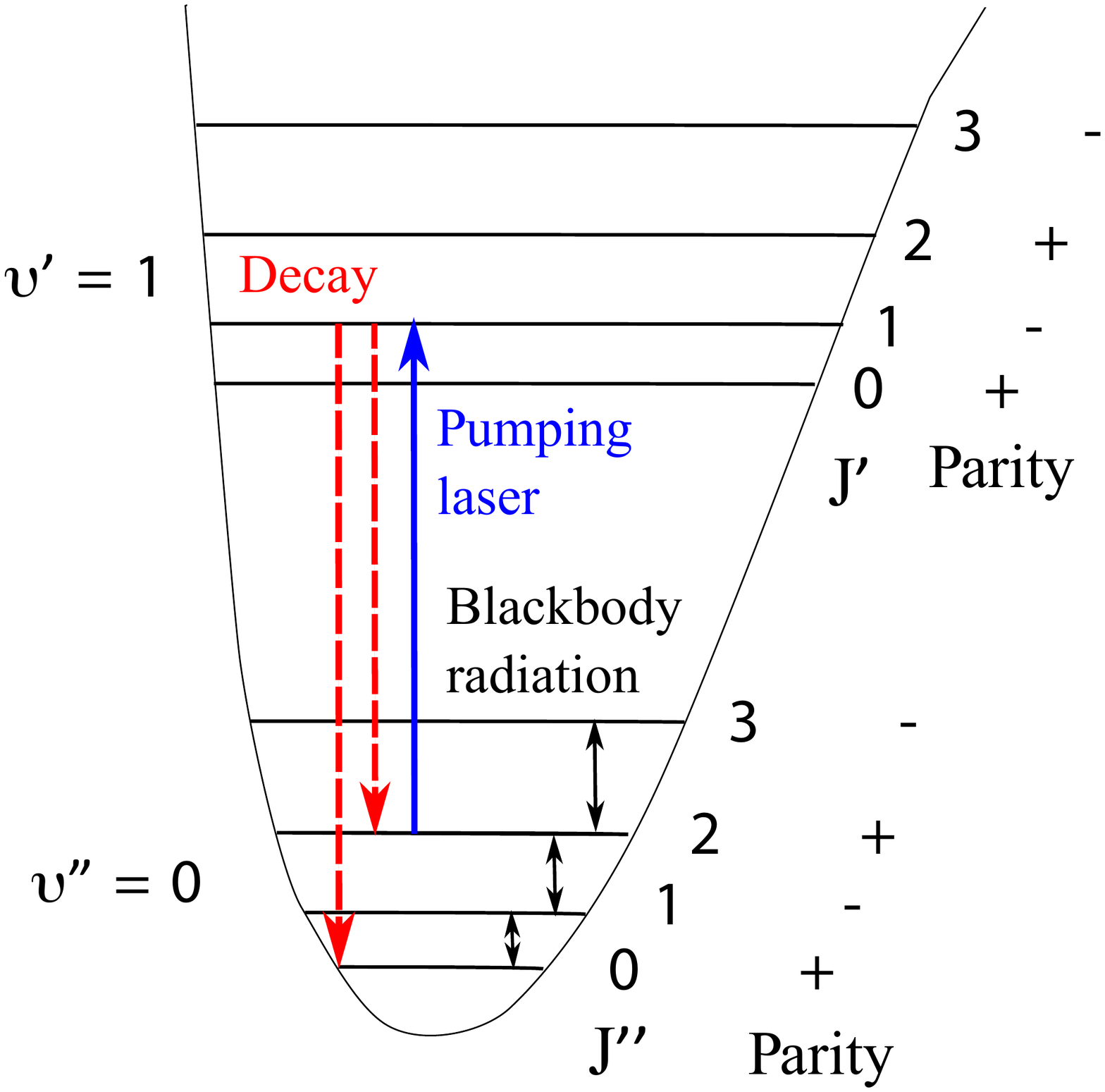}}
    \caption{Schematic of a rotational laser cooling scheme applied for MgH$^{+}$ molecular ions \cite{staanum10a}. The blue solid line represents the laser which pumps population from the $J'' = 2$ rotational state in the vibrational ground state to $J' = 1$ rotational state in the vibrational excited state satisfying $\Delta J = -1$ and $\pm \rightarrow  \mp$ electronic dipole selection rules. The dashed red lines represent spontaneous emission processes which lead to an accumulated ion population in the rotational ground state $J'' = 0$. The black lines indicate black body radiation which transfers population between adjacent rotational levels.
}
    \label{fig:rotationalcooling}
\end{figure}

Thus, in sympathetic cooling experiments state-preparation has to be achieved by additional methods. Based on a proposal by Vogelius et al. \cite{vogelius02a}, laser cooling of the rotational degrees of freedom has been implemented in systems such as MgH$^+$ and HD$^+$ \cite{staanum10a, schneider10a}. The principal idea is to use infrared lasers to optically pump on selected rotational-vibrational transitions in order to achieve an accumulation of the population in a specific quantum state, usually the rotational-vibrational ground state. The scheme is shown in Fig.~\ref{fig:rotationalcooling}. In a typical scheme, population is pumped from selected rotational levels of the ground vibrational state to specific excited states which can decay to the rotational ground state by fluorescence. Initially a range of rotational states are populated at room temperature. However, the use of only one \cite{staanum10a} or two \cite{schneider10a} pumping lasers, each pumping on only one specific rotational level, has been shown to be sufficient to achieve a significant enhancement of the population in the rotational ground state because all levels are coupled by the ambient blackbody-radiation field. Thus, ground-state populations of 37\% in MgH$^+$ with one laser \cite{staanum10a} and 78\% in HD$^+$ with two lasers \cite{schneider10a} have been achieved starting from room-temperature thermal population distributions. Recently, broadband optical pumping on rovibronic transitions, simultaneously addressing a multitude of rotational states, has been implemented using an optical frequency comb \cite{lien14a}. With this approach, it has been possible to reach ground state populations on the order of 95\% after an irradiation time of 140~ms in AlH$^+$. Moreover, sympathetic cooling has recently been combined with cryogenic buffer gas cooling enabling the generation of internally cold molecular ions with translational temperatures down to the millikelvin regime \cite{hansen14a}.

\begin{figure}[!htb]
    \centering
    \resizebox{0.6\linewidth}{!}
    {\includegraphics{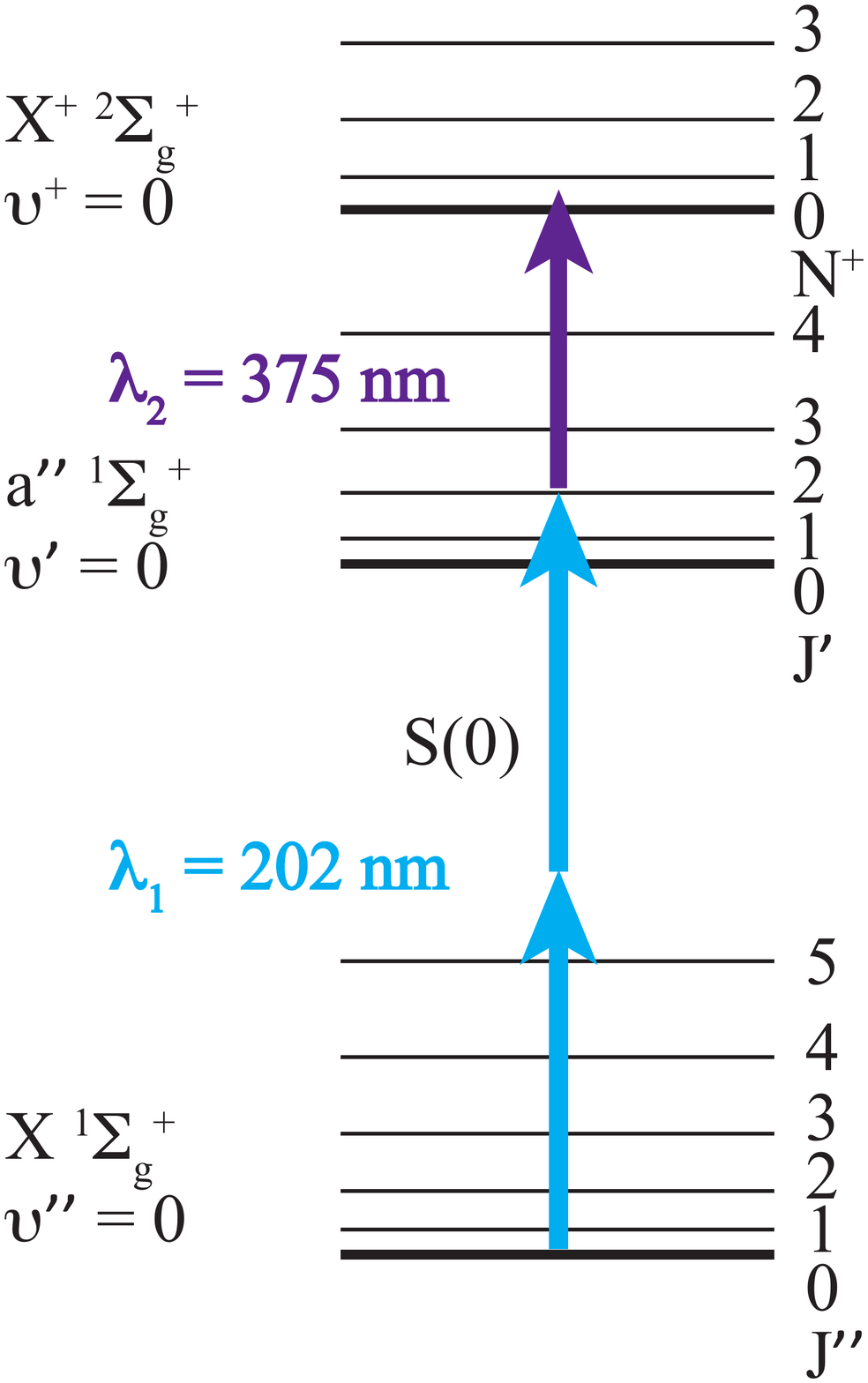}}
    \caption{State-selective [2+1'] resonance-enhanced multi-photoionization (REMPI) scheme for the generation of N$_{2}^{+}$ ions in the $N^{+}$ = 0 rotational states \cite{tong10a,tong11a}. The selected intermediate states are prepared by absorbing two 202 \si{\nano\meter} photons vis $S(0)$ transition from the ground electronic state. The state selection in the ionic state is achieved by threshold photoionization using another 375 \si{\nano\meter} photon.
}
    \label{fig:REMPI}
\end{figure}

An alternative technique to generate molecular ions in specific vibrational-rotational states is resonance-enhanced multi-photon ionization (REMPI), i.e., photoionization of a neutral molecule via a well-defined intermediate state. In these experiments, vibrational state selection of the ion can be achieved by excitation to a selected vibrational level of a suitable Rydberg state of the neutral molecule and taking advantage of diagonal Franck-Condon factors in the subsequent ionization step \cite{morrison86a, anderson92a, sage10a}. In addition, rotational state selection of the ion can be obtained by pulsed-electric-field ionization of high Rydberg states converging on specific rotational ionization thresholds (mass-analyzed threshold ioniozation, MATI) \cite{mackenzie94a, green00a} or by direct ionization slightly above the lowest rotational ionization threshold which is accessible from a specific intermediate state by virtue of photoionization selection and propensity rules (see Fig.~\ref{fig:REMPI}) \cite{willitsch05a}. The latter approach was recently applied to the production of N$_2^+$ ions in specific rotational states which were subsequently sympathetically cooled and Coulomb-crystallized by the interaction with laser-cooled Ca$^+$ ions \cite{tong10a, tong11a}. 

\section{Review of selected results}\label{se:experiment}


\subsection{Temperature dependence of reaction rates}

\label{ss:molbeam}

Both the SIFT and CRESU methods have successfully been employed in a range of measurements on the temperature dependence of ion-molecule reaction rates. A few representative results are discussed in the following.

\begin{figure}[!htb]
    \centering
    \resizebox{0.6\linewidth}{!}
    {\includegraphics{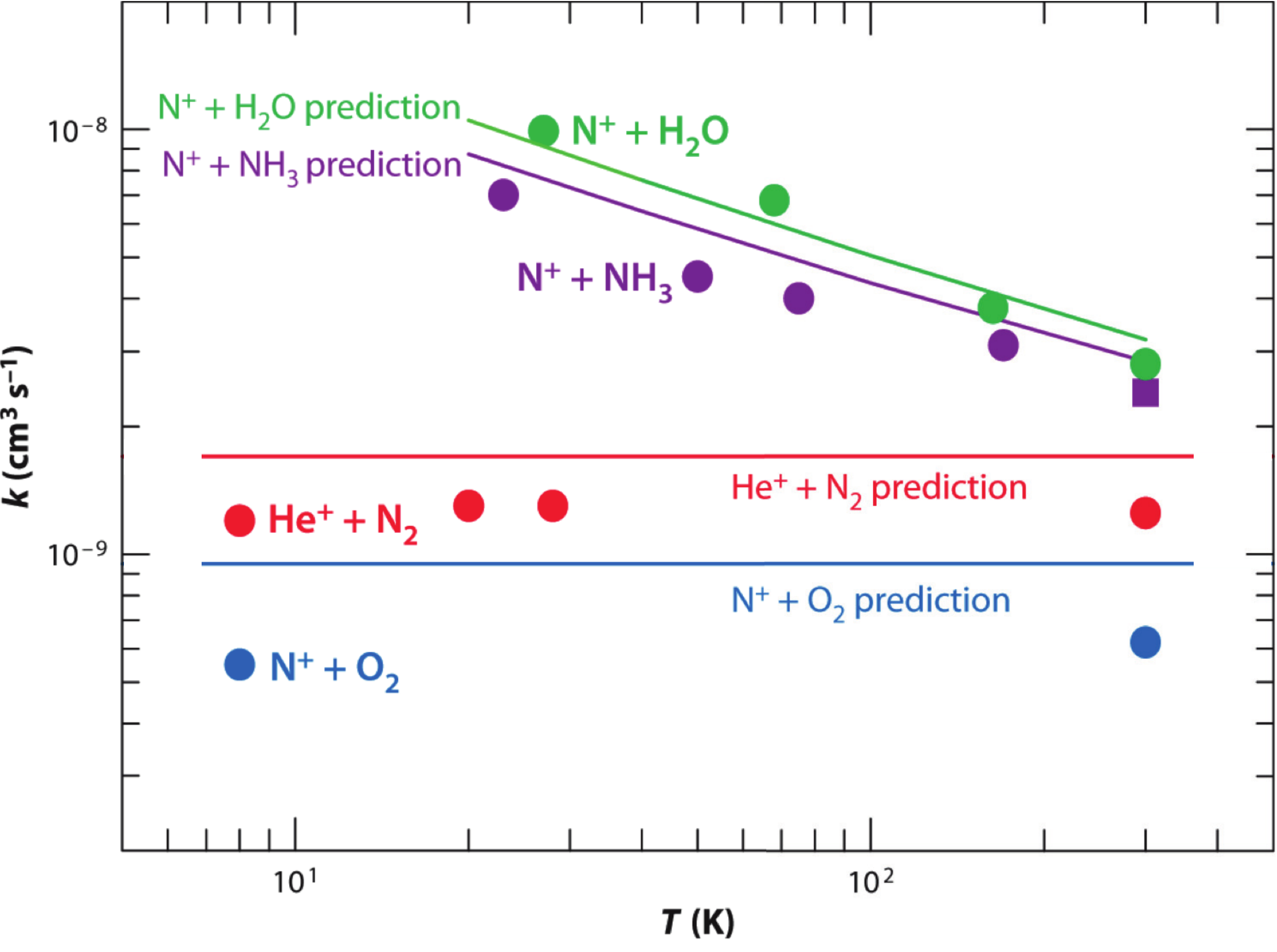}}
    \caption{Examples of rate constants for reactions of ions with polar and non-polar neutral molecules obtained from CRESU measurements. The symbols show experimental
results for N$^{+}$ + H$_{2}$O (green circles), N$^{+}$ + NH$_{3}$ (purple circles and squares), He$^{+}$ + N$_{2}$ (red circles) and N$^{+}$ + O$_{2}$ (blue circles). The lines represent theoretical predictions for the temperature dependence of the relevant rate constants, see text for details. Figure reproduced with permission from Ref. \cite{smith11b}. Copyright 2011 by Annual Reviews.
}
    \label{fig:cresuresults}
\end{figure}

Barrierless reactions between ions and polar molecules which are dominated by the charge-dipole interaction have been predicted to exhibit a pronounced negative temperature dependence of the reaction rate \cite{su73a,clary85b, su82a,chesnavich80a, celli80a,sakimoto80a, sakimoto81a,morgan87a, troe85a,clary87a,troe87a}, see also Table. \ref{tab:classRateSummary}. These predictions have been confirmed in experiments \cite{marquette85b, rebrion88a}. The experimental results were reproduced by adiabatic capture theory and the statistical adiabatic channel model in the temperature range 27--300 \si{\kelvin}  \cite{clary87a,troe87a}. It was also suggested that a simple rate coefficient formula \cite{su82a} is applicable to many fast reactions of ions with polar molecules over this temperature range. From an empirical fit to classical trajectory calculations, the rate constant was found to be \cite{su82a,maergoiz09a}
\begin{equation}
\label{eq:sc1}
k = (0.4767x+0.6200)k_{L},
\end{equation}
if $x \geqslant 2$, where $x = \mu _{D}/\sqrt{2\alpha k_{B}T}$. $ \mu _{D}$ and $\alpha$ are the dipole moment and the polarizability of the neutral reactant, respectively, $k_{B}$ is the Boltzmann constant and $k_{L}$ is the Langevin capture rate constant, ses Sec. \ref{ss:capture}. Note that Eq. (\ref{eq:sc1}) deviates from the simple expressions in Table. \ref{tab:classRateSummary} because it also incorporates the rotation of the dipolar molecules. If $x < 2$, the rate constant was empirically found to be 
\begin{equation}
\label{eq:sc2}
k = ((x + 0.5090)^{2}/10.526 + 0.9754)k_{L}. 
\end{equation}

Fig. \ref{fig:cresuresults} shows rate constants for various reactions between atomic ions and non-polar as well as polar neutral molecules obtained at temperatures T $\leq$ 300 \si{\kelvin} as compiled by Smith \cite{smith11b}. The data have been obtained using the CRESU method. The reaction rates of N$^+$ with the polar neutrals H$_2$O and NH$_3$ show a pronounced dependence on the temperature. The green and purple lines show predictions of the corresponding rate constants using the empirical formulae Eqs. (\ref{eq:sc1}) and (\ref{eq:sc2}), respectively, as discussed above. 

Contrary to reactions with polar molecules, reactions of ions with many non-polar species were found to be temperature independent. Experimental investigations yielded data at temperatures down to 8 \si{\kelvin} \cite{rowe85a} for reactions of He$^{+}$ ions with N$_{2}$, O$_{2}$ and CO, and of N$^{+}$ ions with O$_{2}$, CO and CH$_{4}$. CO has a small permanent dipole moment on the order of 0.1~D and all of these molecules have only moderate quadrupole moments. Fig. \ref{fig:cresuresults} shows the experimental results obtained for He$^+$ + N$_2$ and N$^+$ + O$_2$ and a comparison with the predictions from the Langevin formula \cite{smith11b}. No significant variation of the rate constants with temperature was observed and the experimental rate constants are in fair agreement with the Langevin predictions. These results suggest that the Langevin formula adequately describes the reaction rates in the temperature interval studied and that corrections introduced by quadrupolar interactions and anisotropic polarizabilities (see for instance the chapter by Lepers and Dulieu in this book) which are neglected in the Langevin picture are not significant for these cases in the temperature range studied.

Three-body association reactions are a class of ionic processes of particular importance in environments with moderate or high pressure. In these reactions, an ion A$^+$ and a neutral species B form an adduct AB$^+$ which is stabilized by collisions with a third body, often an inert-gas molecule. The temperature dependences of three-body ion-molecule association reactions such as CH$_{3}^{+}$ + H$_{2}$, CO, N$_{2}$, O$_{2}$ CO$_{2}$; N$_2^+$ + N$_2$; O$_2^+$ + O$_2$; Xe$^{+}$ + Ar; He$^{+}$ + H$_2$; He$^+$ + He; Ne$^{+}$ + He; Ar$^{+}$ + Ar; Kr$^{+}$ + Ar; O$_{2}^{+}$ + Kr,  NO$^{+}$ + Kr, O$_{2}^{+}$ + Ar \cite{adams79a,adams81a,jones80a,jones80b,bohringer82a,bohringer83a,bohringer86a,smith84a, ferguson84a} 
have been measured  in various inert gases using the SIFT method. The rate constants are frequently be described by power-law expressions of the form 
\begin{equation}\label{eq:kN}
k=AT^{-n},
\end{equation}
which were often found to be valid over large temperature ranges. 

Apart from studies of barrierless processes which can often adequately be understood within the framework of capture models, the experimental methods discussed here have also been used to characterize reactions with activiation barriers. One exemplary CRESU study focused on the slightly endothermic reaction
\begin{equation}\label{eq:Hreac}
N^{+} (^{3}P) + (para-, ortho-)H_{2} \rightarrow NH^{+} + H
\end{equation}
at temperatures below 163~K \cite{marquette88a}. It was found that the rate constant markedly decreased with decreasing temperature indicating that the reaction requires activation energy. Moreover, it was found that the rate was enhanced when using ortho-, i.e., rotationally excited, H$_{2}$, indicating the importance of rotational energy for promoting these reactions. 

Cryogenic ion traps have recently been established as an alternative to the SIFT and CRESU methods for studies of ion-molecule reactions at low temperatures, see Sec. \ref{ss:buffergas}. A range of ion-molecule processes, many of which are important for astrophysics, have recently been studied with this approach. They include deuteration reactions of small hydrocarbon ions such as CH$_n^+~(n=3-5)$ and C$_x$H$_y^+$ \cite{gerlich02a, asvany04b, asvany05b}, reactions of CH$_x^+$ and CO$_2^+$ with H and H$_2$ \cite{gerlich11a, borodi09a}, and laser-induced charge-transfer reactions, e.g., N$_2^+$ + Ar \cite{schlemmer99a}.

One key advantage of the ion-trap technique lies in the possibility to perform reactions experiments over considerably longer time scales than is possible with the SIFT or CRESU methods, which enables the measurement rate constants of very slow reactions. Examples include radiative association reactions such as CH$_3^{+}$ + H$_{2}$ \cite{gerlich95a}. Another example is the competition between ternary and radiative association reactions, e.g.,
\begin{equation}
\label{eq:iontrapreaction1}
CH_{3}^{+} + H_{2}O + He\rightarrow CH_{5}O^{+} + He,
\end{equation}
\begin{equation}
\label{eq:iontrapreaction2}
CH_{3}^{+} + H_{2}O \rightarrow CH_{5}O^{+} + h\nu,
\end{equation}
which has been studied in an ion trap at 20~K \cite{gerlich06a}.

Apart from reactions with cations, processes with anions have also been investigated using cryogenic traps. These include electron transfer and associative detachment of H$^-$/D$^-$ with H \cite{rouvcka12a,rouvcka15a} and cold reactions of NH$_2^-$ with H$_2$ which were found to exhibit an unusual low-temperature behavior \cite{otto08a}. Moreover, state-resolved inelastic collisions of cold OH$^-$ anions with He have recently also been investigated in a cryogenic trap \cite{hauser15a}.

\subsection{Combined cold-molecules cold-ions experiments}\label{ss:hybridtrap}

\begin{figure}[!htb]
    \centering
    \resizebox{\linewidth}{!}
    {\includegraphics{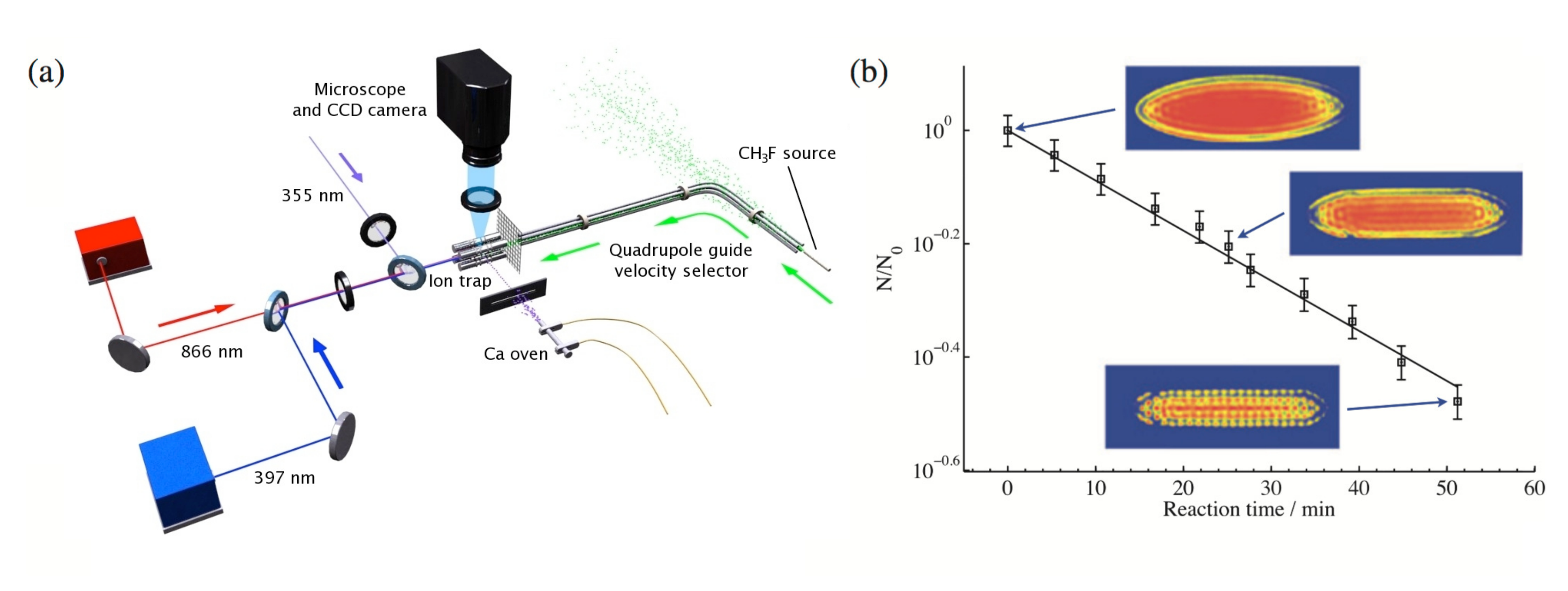}}
    \caption{(a) Experimental setup for studying cold ion-molecule reactions. A linear RF trap for the generation of Coulomb crystals of cold ions is combined with an electrostatic velocity selector for polar neutral molecules \cite{willitsch08a}. (b) Fraction of laser-cooled Ca$^{+}$ ions remaining in the Coulomb crystal as a function of the time of reaction with velocity-selected CH$_{3}$F molecules. The solid line represents a fit to an integrated pseudo-first-order rate law from which bimolecular rate constants for the reaction Ca$^{+}$ + CH$_{3}$F $\rightarrow$ CaF$^{+}$ + CH$_{3}$ were obtained. The insets show false-color fluorescence images of the Coulomb crystal over the course of the reaction. Figure adapted from Ref. \cite{willitsch12a}: Coulomb-crystallised molecular ions in traps: methods, applications, prospects, Stefan Willitsch, International Reviews in Physical Chemistry, published on 27 Mar 2012. Reprinted by permission of Taylor \& Francis.
    }
    \label{fig:ccrxn}
\end{figure}

The flow, beam and buffer-gas cooling experiments discussed so far are all limited to temperatures in the range $\geqslant 10$~K. Studies at lower temperatures require alternative approaches. For that purpose, experimental setups have recently been developed which combine methods for the generation of both cold neutrals and cold ions.

One of the early experiments in this domain consisted of the combination of a linear RF quadrupole ion trap for the generation of Coulomb crystals with a velocity selector for the preparation of cold neutral molecules \cite{willitsch08a}. A schematic of the experimental setup is presented in Fig. \ref{fig:ccrxn}. The velocity selector consists of a bent electric quadrupole guide into which a thermal beam of polar molecules is injected \cite{rangwala03a}. Molecules in low-field-seeking Stark states (which are pushed from the high field regime to the low field regime) are retained in the guide if their transversal kinetic energy does not exceed their Stark energy at the edges of the quadrupolar electric field generated by the arrangement of electrodes. When the guided molecules reach the bend, only the slowest molecules whose Stark energy exceeds the centrifugal energy in the bend are forced around the corner, whereas the fast molecules are ejected from the guide. Thus, a beam of translationally cold molecules is produced at the exit of the quadrupole where the ion trap is placed. The translational temperature of the polar molecules is typically in the Kelvin range, whereas the kinetic energy of the trapped ions can be close to the Doppler-cooling limit for strings of ions located on the central trap axis where the micromotion vanishes. 

Processes studied with this approach so far include reactions of laser-cooled Ca$^+$ ions with velocity-selected CH$_3$F, CH$_2$F$_2$ and CH$_3$Cl \cite{willitsch08a, gingell10a}. In these systems, the reaction rates are not capture limited, but governed by electronic barriers at short range. Depending on the specific system, the barriers are either calculated to be slightly submerged, i.e., their maximum is located below the energy of the reactants, as in CH$_3$F and CH$_3$Cl, or they are true barriers which slow down the reactions at low energies, as in CH$_2$CF$_2$ \cite{gingell10a}. In the latter case, the reaction essentially only occurs with Ca$^+$ ions which are electronically excited to the $(4p)~^2P_{1/2}$ and $(3d)~^2D_{3/2}$ states during laser cooling.  

The method has subsequently been extended to reactions with sympathetically cooled ions, e.g., OCS$^+$ + ND$_3\rightarrow~$ND$_3^+$ + OCS \cite{bell09a}. Similar experiments have recently also been reported on Ca$^+$ + ND$_3$, Ca$^+$+CH$_3$CN and N$_2$H$^+$ + CH$_3$CN \cite{okada13a}.

\subsection{Ion-atom hybrid traps}\label{ss:atomhybridtrap}

\begin{figure}[!htb]
    \centering
    \resizebox{0.6\linewidth}{!}
    {\includegraphics{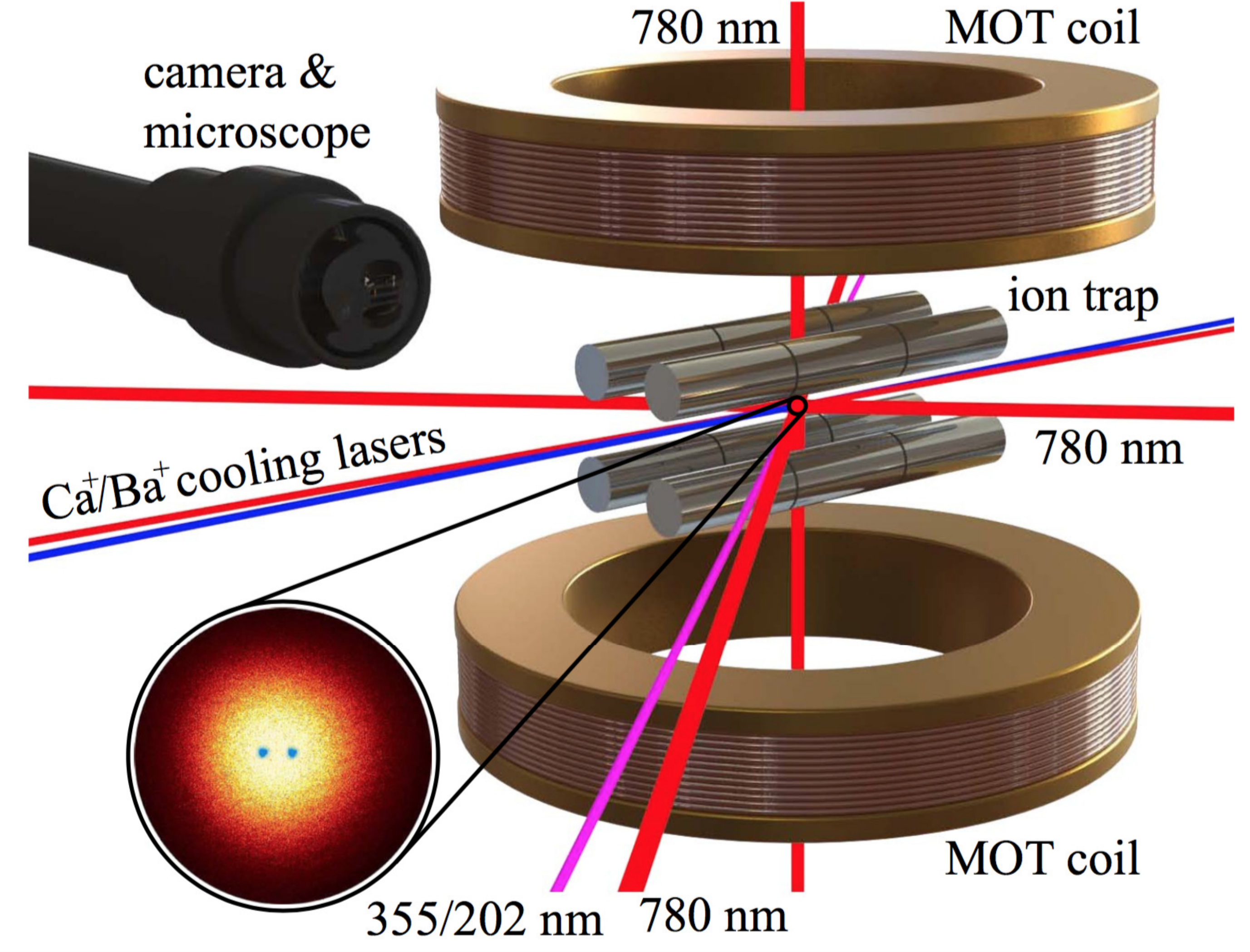}}
    \caption{Schematic of a hybrid trapping experiments consisting of a magneto-optical trap (MOT) for ultracold atoms and a linear quadrupole RF ion trap. Laser beams at 780~nm serve for the laser cooling of Rb atoms, the other laser beams are used for the generation and laser cooling of ions. The two coils generate a quadrupolar magnetic field required for operation of the MOT. The inset shows a false-color fluoresence image of two laser-cooled Ca$^+$ ions (blue) immersed in cloud of ultracold Rb atoms (yellow and red). Figure adapted from Ref. \cite{hall13a}: Ion-neutral chemistry at ultralow energies: dynamics of reactive collisions between laser-cooled Ca$^{+}$ ions and Rb atoms in an ion-atom hybrid trap, Felix H.J. Hall, Pascal Eberle, Gregor Hegi, Maurice Raoult, Mireille Aymar, Olivier Dulieu and Stefan Willitsch, Molecular Physics, published on 08 Apr 2013. Reprinted by permission of Taylor \& Francis.
    }
    \label{fig:hyb}
\end{figure}

The lowest energies which can currently be reached in ion-neutral collisions are achieved in ion-atom hybrid trapping experiments \cite{sias14a, haerter14a,willitsch15a}. In these experiments, cold trapped ions are combined with ultracold trapped atoms enabling collisions experiments in the millikelvin regime. In most setups of this type, a magneto-optical trap (MOT) for neutral atoms is combined with a quadrupole RF ion trap,  see Fig. \ref{fig:hyb} \cite{smith05a,grier09a, hall11a, rellergert11a, ravi11a, haze13a}. In a MOT \cite{raab87a}, atoms are cooled and trapped by the combined action of six laser beams for laser cooling (two counterpropagating beams along each coordinate axis) with a quadrupolar magnetic field. Refer to, e.g., Refs. \cite{metcalf99a, foot05a} for a detailled discussion of the working principle of a MOT. Another class of experiments relies on the combination of a magnetic trap \cite{zipkes10a} or an optical dipole trap \cite{schmid10a, meir16b} for the ultracold atoms with an RF ion trap. In these experiments, the neutral atoms can be cooled down to several hundreds of nanokelvin enabling the generation of quantum-degenerate gases \cite{zipkes10a}. 

So far, collisions and reactions of a variety of ions such as Na$^{+}$, Yb$^{+}$, Rb$^{+}$, Ba$^{+}$, N$_{2}^{+}$, BaCl$^{+}$ with neutral atoms such as Na, Rb and Ca have been studied \cite{sias14a, haerter14a, willitsch15a}. In general, these experiments confirmed that the reaction rates are bounded from above by the relevant Langevin rates (see Table~\ref{tab:classRateSummary}). While early studies focused on the sympathetic cooling of the translational motion of the ions by the ultracold atoms \cite{zipkes10a, schmid10a, sivarajah12a, haze13a}, the focus has shifted to the study of chemical reactions in recent years. While radiative and non-adiabatic charge exchange has been shown to be an important reactive process in these systems \cite{grier09a, zipkes10b, schmid10a, hall11a, rellergert11a}, there is also a growing body of experimental and theoretical evidence for the generation of molecular ions by radiative association \cite{hall11a, hall13a, hall13b, dasilva15a, sayfutyarova13a} and potentially three-body collisions \cite{haerter12a}. The reaction rates were also shown to be considerably enhanced by electronic excitation of the ions in a variety of systems \cite{hall11a, rellergert11a, ratschbacher12a, sullivan12a, hall13a, hall13b}.   
Morover, evidence for the cooling of the vibrational degrees of freedom of BaCl$^+$ ions by collisions with ultracold Ca atoms has recently been found \cite{rellergert13a}.

\begin{figure}[!htb]
    \centering
    \resizebox{\linewidth}{!}
    {\includegraphics{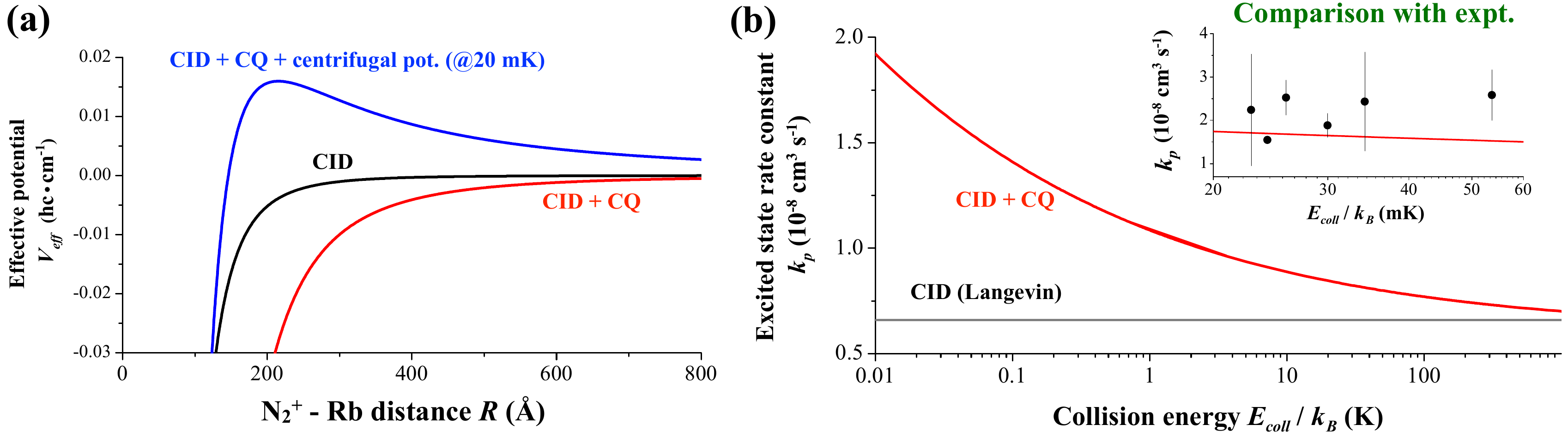}}
    \caption{Cold reactions of sympathetically cooled N$_2^+$ ions with laser-cooled Rb atoms in an ion-atom hybrid trap. (a) Long-range potentials accounting for the charge-induced dipole (i.e., Langevin) interaction (CID, black line) and the CID plus charge-quadrupole (CQ) interactions (red line). The blue line shows the centrifugally corrected potential at a collision energy corresponding to 20~mK. (b) Classical capture rate constants for CID (black line) and CID+CQ potentials as a function of the collision energy. The inset shows a comparison with the experimental results in a collision energy interval corresponding to 20--60~mK. Figure reproduced from Ref. \cite{willitsch15a}.}
    \label{fig:n2+rb}
\end{figure}

Reactions at very low temperatures are a useful tool to probe fine details of intermolecular interactions which are often not manifest in experiments at higher temperatures. One example are reactions of sympathetically cooled N$_2^+$ ions with ultracold Rb atoms which were excited to the $(5p)~^2P_{3/2}$ state in a hybrid trap \cite{hall12a}. Rate constants for charge-transfer reactions were probed in an interval of collision energies corresponding to 20--60~mK and were found to exceed the predictions from classical Langevin capture theory (Sec. \ref{ss:capture}) by about a factor of 4 in this regime. However, the experimental data could well be reproduced by a classical capture model including the interaction between the charge of the ion and the quadrupole moment of Rb in its excited $(5p)~^2P_{3/2}$ state, see Fig. \ref{fig:n2+rb}. These findings suggest that the capture at the low collision energies achieved in this study is dominated by the charge-quadrupole interaction. Fig. \ref{fig:n2+rb} (b) also shows that the effect of the charge-quadrupole interaction on the rate constants diminishes with increasing energy because of the negative temperature dependence of the corresponding capture rate constant, see Sec. \ref{ss:capture}. Thus, at collision energies of the order of tens to hundreds of Kelvins, the rate constants are expected to approach the Langevin limit. This finding is in line with the results of the CRESU measurements on reactions of ions with non-polar molecules discussed in Sec. \ref{ss:molbeam} which all showed rate constants close to the Langevin limit in this regime.

One of the frontiers of cold-collision studies are experiments on single, state-selected particles. In this context, Ratschbacher et al. \cite{ratschbacher12a} reported results on single Yb$^+$ ions immersed into a cloud of state-selected ultracold Rb atoms. The Yb$^+$ ions were  prepared in selected quantum states in order to study fully state-specific reaction rates with Rb. Besides an acceleration of the reaction rate upon electronic excitation of the ion, the reaction rate was found to sensitively depend on the hyperfine state of the Rb atoms, underlining the importance of subtle effects like hyperfine interactions in cold collisions \cite{Tscherbul16a}.

\section{Conclusions and outlook}\label{se:conclusion}

Studies of ion-neutral reactions at low temperatures have made impressive progress over the recent years. A range of beam, flow and trap techniques now enables to study ion-neutral processes down to the millikelvin regime. Besides yielding important kinetic data for modelling the chemistry of cold and dilute environments such as interstellar space, studies in this regime have revealed "exotic" chemical processes such radiative association and allowed to probe the effects of long-range intermolecular forces on the reaction kinetics. 

\begin{figure}[!htb]
    \centering
    \resizebox{0.8\linewidth}{!}
    {\includegraphics{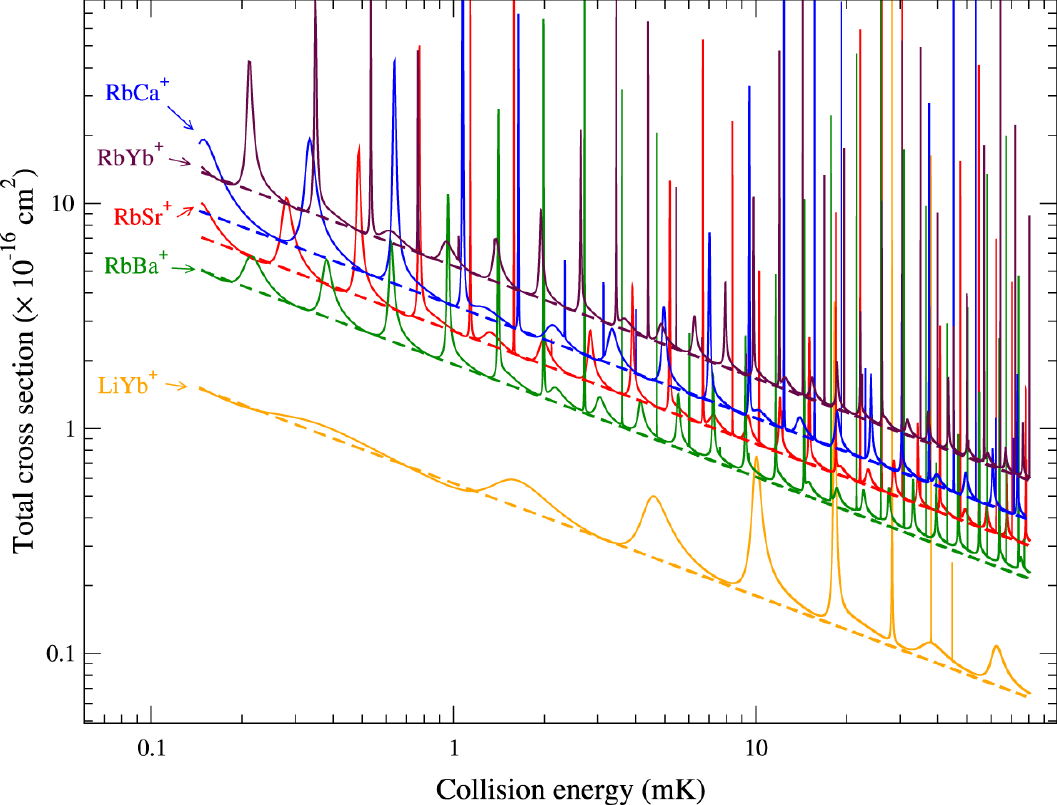}}
    \caption{Calculated total cross sections (RA + RCT) from Eqs.~(\ref{csrct}) and ~(\ref{csra}), as functions of the collision energy (with both axes scaling logarithmically). The dashed lines correspond to the classical cross section obtained from Langevin theory. Figure adapted from Ref. \cite{dasilva15a}: Humberto da Silva Jr, Maurice Raoult, Mireille Aymar and Olivier Dulieu, Formation of molecular ions by radiative association of cold trapped atoms and ions, Humberto da Silva Jr et al 2015 New J. Phys. 17 045015, DOI: http://dx.doi.org/10.1088/1367-2630/17/4/045015.
    }
    \label{fig:resonances}
\end{figure} 

A future challenge, from the perspective of 2016, will be to reach even lower collision energies to access a regime in which collisions are dominated by only a few partial waves and quantum effects strongly influence the collision dynamics. An example is presented in Fig. \ref{fig:resonances} which shows theoretical predictions of cross sections for radiative association and radiative charge transfer in cold collisions of Ca$^+$ with Rb. The energy spectrum of the cross sections shows sharp features corresponding to shape resonances at which the collision energy matches the energy of a metastable state trapped behind the centrifugal barrier. The position and widths of these resonances serve as a very sensitive probe of the collisional angular momentum and the details of the interaction potential \cite{henson12a}. The observation of these elusive quantum features will require new experiments with a better energy tunability at an improved energy resolution in the sub-Kelvin regime. 

A further objective will be the improvement of translational-cooling methods for molecular ions. Lower translational temperatures may be achievable by laser cooling of selected molecular ions with a sufficiently simple energy-level structure \cite{nguyen11a}, as has recently been demonstrated for neutral molecules \cite{shuman10a}. Moreover, at low energies fine details of the molecular energy level structure can have pronounced influences on the collision dynamics. Thus, it is desirable to improve present state-preparation techniques and also achieve control over the fine and hyperfine degrees of freedom of molecular ions \cite{bressel12a}. In addition, methods need to be developed which allow an improved characterization of the state distributions of the reaction products with the ultimate aim to extract detailed state-to-state reaction cross sections. Finally, new types of experiments will open up new perspectives for highly controlled ion-molecule collision and reaction experiments. These include the integration of buffer-gas cooling and sympathetic cooling \cite{hansen14a} as well as the combination of Coulomb-crystal experiments with advanced cold-molecules sources such as Stark and Zeeman decelerators \cite{bell09b, heazlewood15a}.

\section*{Acknowledgements}

We thank Prof. D. Gerlich for providing the figure of his setup. This work is supported by the Swiss National Science Foundation, grant nrs. BSCGI0\_157874 and 200021\_156182, and the European Commission under the Seventh Framework Programme FP7 GA 607491 COMIQ.

\bibliography{ColdIonChemistry,Main-Mai2016}
\end{document}